\documentclass{article}

\title{Renormalization of lattice-regularized quantum gravity models \\II. The case of causal dynamical triangulations}

\author{Joshua H. Cooperman \\ \emph{Institute for Mathematics, Astrophysics, and Particle Physics}\\ \emph{Radboud Universiteit Nijmegen, Heyendaalseweg 135, 6526 AJ Nijmegen, Nederland}}

\usepackage{float}
\usepackage{multicol}
\usepackage{subfigure}
\usepackage{tabularx}
\usepackage{booktabs}
\usepackage{graphicx}
\usepackage{amsmath}
\usepackage{amsfonts}
\usepackage{latexsym}
\usepackage{mathrsfs}
\usepackage{geometry}
\usepackage{chngcntr}
\counterwithin{figure}{section}
\counterwithin{table}{section}
\numberwithin{equation}{section}

\begin{document}

\maketitle

\begin{abstract}
The causal dynamical triangulations approach aims to construct a quantum theory of gravity as the continuum limit of a lattice-regularized model of dynamical geometry. A renormalization group scheme---in concert with finite size scaling analysis---is essential to this aim. Formulating and implementing such a scheme in the present context raises novel and notable conceptual and technical problems. 
I explored these problems, and, building on 
standard techniques, suggested potential solutions in the first paper of this two-part series. 
As an application of these solutions, I now propose a renormalization group scheme for causal dynamical triangulations. This scheme differs significantly from that studied recently by Ambj\o rn, G\"{o}rlich, Jurkiewicz, Kreienbuehl, and Loll. 
\end{abstract}

\section{Introduction}\label{introduction}

The causal dynamical triangulations approach to the construction of quantum theories of gravity is patterned on lattice regularization approaches to the construction of quantum theories of fields.  
While one introduces a lattice regularization to render the quantum theory well-defined, one is ultimately interested in the hypothetical continuum limit in which the regularization is removed in such a manner that physical quantities remain finite. The existence of a continuum limit of the lattice-regularized quantum theory is typically contingent on this theory exhibiting a second order phase transition. The nontriviality of a continuum limit is then contingent on the existence of an ultraviolet fixed point 
along the second order phase transition. The identification of such a fixed point calls for a renormalization group analysis, which one performs in concert with finite size scaling. 

There is now compelling evidence for the existence of a second order phase transition within the causal dynamical triangulations of $(3+1)$-dimensional Einstein gravity \cite{JA&SJ&JJ&RL1,JA&SJ&JJ&RL2}. This second order phase transition moreover occurs at the boundary of the phase exhibiting physical quantum geometry, potentially implying that the hypothetical continuum limit also possesses this property. A search for this continuum limit in the form of an ultraviolet fixed point is thus clearly warranted. Ambj\o rn, G\"{o}rlich, Jurkiewicz, Kreienbuehl, and Loll very recently made the first such attempt \cite{JA&AG&JJ&AK&RL}. Under their simplest interpretation of the continuum limit, they did not find evidence for the existence of an ultraviolet fixed point, but, under a more complicated interpretation of the continuum limit, they did find plausible evidence for the existence of an ultraviolet fixed point. I propose a renormalization group scheme that differs significantly from those of these authors. Hopefully, my proposal assists in the search for an ultraviolet fixed point within the causal dynamical triangulations of $(3+1)$-dimensional Einstein gravity. 

\subsection{Recapitulation of paper I}\label{recap}

In paper I of this two-part series, I considered the formulation and implementation of a renormalization group scheme for lattice-regularized quantum theories of gravity \cite{PaperI}. Drawing 
on the well-established methods for renormalization group analysis of lattice-regularized quantum theories of fields, 
I proposed a general method for renormalization group analysis of lattice-regularized quantum theories of gravity. I now review my proposal, dividing its procedure into six steps. The first four steps, although separated in the following for clarify of presentation, are so intimately connected that one typically pursues them all at once. Moreover, one should subject these four steps to a comprehensive statistical analysis, only proceeding subsequently with the last two steps.

\subsubsection{Selection of a model for the continuum limit}

One first selects a model for the continuum limit of the lattice-regularized quantum theory of gravity. A model consists of a progression of theories $\mathscr{T}$ along a renormalization group trajectory within some truncation $\mathfrak{t}$ of the space $\mathfrak{T}$ of all theories. From numerical measurements interpreted with respect to the model, one aims to specify this progression of theories by determining the values of the renormalized couplings characterizing each theory. One constructs the renormalization group flows of the model's couplings in this manner. 

There is of course no guarantee that the lattice-regularized quantum theory of gravity possesses a continuum limit. Typically, the existence of a continuum limit requires the existence of a second order phase transition within the quantum theory's phase structure. Even if a continuum limit does not exist, the theory might prove effective over a considerable range of scales, so one might still want a continuous description with a finite ultraviolet cutoff. 

\subsubsection{Selection of a finite size scaling \emph{Ansatz}}

One next selects a finite size scaling \emph{Ansatz} relating discrete quantities within the lattice-regularized quantum theory of gravity to continuous quantities within the chosen model for its continuum limit. This \emph{Ansatz} should be consistent with the chosen model in that it should be based on a correspondence between a physical observable of the chosen model and the hypothesized discrete analogue of this physical observable. 
The form of the finite size scaling \emph{Ansatz} may vary across the phase diagram of the lattice-regularized quantum theory of gravity as different scaling regimes are encountered. 

\subsubsection{Delineation of renormalization group trajectories}

One delineates renormalization group trajectories as follows. 
Consider a physical observable $\mathscr{O}(\ell)$ of the chosen model for the continuum limit whose value generically depends on the length scale $\ell$ which one probes.\footnote{Whenever I refer to a scale, I always refer to a length scale.} Suppose that one has experimental access to $\mathscr{O}(\ell)$ on the interval of scales $(\ell_{\mathrm{UV}},\ell_{\mathrm{IR}})$. A renormalization group transformation effects a change in the interval of scales to which one has access, in general, from $(\ell_{\mathrm{UV}},\ell_{\mathrm{IR}})$ to $(\ell'_{\mathrm{UV}},\ell'_{\mathrm{IR}})$. Since a renormalization group transformation does not effect changes in physics, measurements of $\mathscr{O}(\ell)$ on the interval of scales $(\ell_{\mathrm{UV}},\ell_{\mathrm{IR}})$ and on the interval of scales $(\ell'_{\mathrm{UV}},\ell'_{\mathrm{IR}})$ must agree within the intersection $(\ell_{\mathrm{UV}},\ell_{\mathrm{IR}})\cap(\ell'_{\mathrm{UV}},\ell'_{\mathrm{IR}})$ of these two intervals. Accordingly, one delineates a renormalization group trajectory as the progression of theories through the space $\mathfrak{T}$ along which the physics of physical observables do not change. The number of physical observables required to delineate a renormalization group trajectory depends on the chosen model. 

A physical observable $\mathscr{O}(\ell)$ is either dimensionless or dimensionful. The value of a dimensionless physical observable at some scale $\ell$ is simply a real number. The value of a dimensionful physical observable at some scale $\ell$ is also a real number, namely the ratio of the physical observable to a standard unit of measure.\footnote{This is what it means for a quantity to be dimensionful: one requires a standard unit of measure to ascertain its value.} This reasoning also applies to the scale $\ell$ at which a physical observable is measured. To measure a dimensionful physical observable, one thus requires an appropriate standard unit of measure. I assume the presence of certain constants of nature, such as the speed of light and the Planck constant, that allow for any physical observable's dimension to be expressed in powers of length. 


Typically, one employs a standard unit of measure defined externally, meaning that the dynamics through which the standard unit of measure is defined are sufficiently decoupled from the dynamics to which the standard unit of measure is applied. In the context of lattice-regularized quantum theories of gravity, external reference scales are usually lacking, so one must choose a dynamically generated scale---itself defined by a particular physical observable---to serve as the standard unit of measure. When one implements a renormalization group scheme, one attempts to ascertain whether or not one's chosen standard unit of measure functions as a good standard. 




When one extracts any dimensionful physical observable from numerical measurements, one necessarily obtains its value in units of the lattice spacing $a$. (The lattice spacing is the only dimensionful quantity present.) For the chosen standard unit of length $\ell_{\mathrm{unit}}$, numerical measurements yield $\tilde{\ell}_{\mathrm{unit}}=\ell_{\mathrm{unit}}/a$. 
Since the lattice spacing is \emph{a priori} arbitrary, one sets its value in units of $\ell_{unit}$. One takes $\tilde{\ell}_{\mathrm{unit}}$ not as a measurement of $\ell_{\mathrm{unit}}$ in units of $a$ but as a definition of $a$ in units of $\ell_{\mathrm{unit}}$. One then measures any other scale $\ell$, determined from numerical measurements only as $\tilde{\ell}=\ell/a$, by forming the ratio $(\ell/a)/(\ell_{\mathrm{unit}}/a)$, which is independent of $a$. In particular, the choice of a standard unit of length $\ell_{\mathrm{unit}}$ makes definite the interval of scales $(\ell_{\mathrm{UV}},\ell_{\mathrm{IR}})$ accessible at some point along a renormalization group trajectory. One defines $\ell_{\mathrm{UV}}$ as the smallest scale present---the lattice spacing $a$---measured in units of $\ell_{\mathrm{unit}}$, and one defines $\ell_{\mathrm{IR}}$ as the largest scale present measured in units of $\ell_{\mathrm{unit}}$. 

\subsubsection{Extraction of the renormalized couplings}

One now determines how to extract the renormalized couplings of the chosen model for the continuum limit. One must perform a set of numerical measurements whose outcomes yield the values of the renormalized couplings when put in correspondence with the chosen model \emph{via} the finite size scaling \emph{Ansatz}. 

\subsubsection{Implementation of a renormalization group transformation}

One next implements a renormalization group transformation. No matter the means by which one effects a renormalization group transformation---typically a coarse graining procedure---it must integrate out degrees of freedom on the interval of scales $(\ell_{\mathrm{UV}},\ell_{\mathrm{IR}})\backslash(\ell'_{\mathrm{UV}},\ell'_{\mathrm{IR}})$ while preserving the physics of physical observables. 

\subsubsection{Construction of the renormalization group flows}

One finally constructs the renormalization group flows of the chosen model's couplings by iterating the renormalization group transformation and extracting their values after each iteration. For a renormalization group transformation that incrementally increases $\ell_{\mathrm{UV}}$ and leaves fixed $\ell_{\mathrm{IR}}$, one may take the value of $\tilde{\ell}_{\mathrm{UV}}$ as the parameter along the flow. 

\subsection{Overview of paper II}

Now, in paper II of this two-part series, I apply this method to the case of causal dynamical triangulations. I review the formalism of causal dynamical triangulations in section \ref{theory}, and I review the phenomenology of causal dynamical triangulations in section \ref{phenomenology}. 
As I argued in paper I, the phenomenology of a lattice-regularized quantum theory of gravity forms the basis of a renormalization group scheme. Accordingly, I provide a comprehensive discussion, paying particular attention to the phenomenological scales that arise dynamically. 
Drawing on this phenomenology, I propose a renormalization group scheme for causal dynamical triangulations in section \ref{renormalization}. This proposal constitutes the paper's primary novel contribution. There remains an ambiguity in my scheme---the choice of a standard unit of length, more than one of which appears to be consistent---that I hope its implementation resolves. 
I conclude 
in section \ref{conclusion} by considering how the observed phase structure of causal dynamical triangulations might mesh with my proposed renormalization group scheme.

\section{The formalism of causal dynamical triangulations}\label{theory}

Causal dynamical triangulations is an approach to the quantization of classical metric theories of gravity based on a particular lattice regularization of the formal path integral
\begin{equation}\label{gravitypathintegral}
\mathscr{A}[\gamma]=\int_{\mathbf{g}|_{\partial\mathscr{M}}=\gamma}\mathrm{d}\mu(\mathbf{g})\,e^{iS_{\mathrm{cl}}[\mathbf{g}]/\hbar}.
\end{equation}
The transition amplitude $\mathscr{A}[\gamma]$ in the quantum theory is specified by the metric tensor $\gamma$ induced on the boundary $\partial\mathscr{M}$ of the spacetime manifold $\mathscr{M}$ by its own metric tensor $\mathbf{g}$. One computes $\mathscr{A}[\gamma]$ by integrating over all metric tensors $\mathbf{g}$ satisfying the boundary condition $\mathbf{g}|_{\partial\mathscr{M}}=\gamma$, weighting each metric tensor $\mathbf{g}$ by the product of the measure $\mathrm{d}\mu(\mathbf{g})$ and the exponential $e^{iS_{\mathrm{cl}}[\mathbf{g}]/\hbar}$.\footnote{I choose to display factors of $\hbar$, but I treat $\hbar$ as having the value unity.} $S_{\mathrm{cl}}[\mathbf{g}]$ is the action of the classical theory of gravity that one wishes to quantize. This computation is, however, plagued by severe conceptual and technical difficulties. What precisely is the class of all metric tensors? Which measure on this class should one employ? How should one regularize the integration's divergences? 

\begin{figure}[ht]
\centering
\subfigure[ ]{
\includegraphics[scale=0.8]{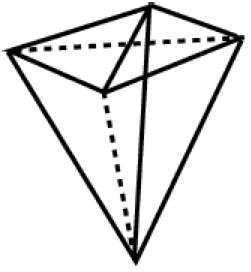}
\label{14simplex}
}
\subfigure[ ]{
\includegraphics[scale=0.8]{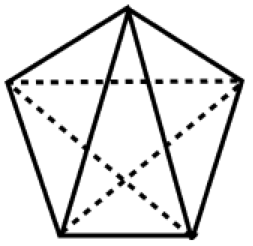}
\label{23simplex}
}
\subfigure[ ]{
\includegraphics[scale=0.8]{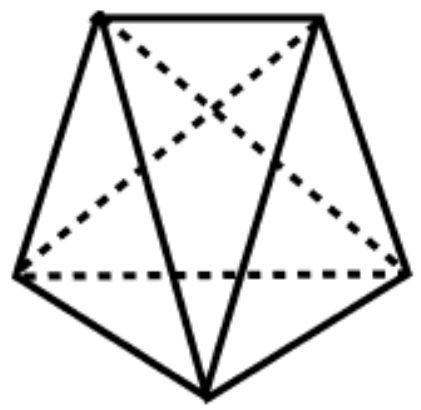}
\label{32simplex}
}
\subfigure[ ]{
\includegraphics[scale=0.8]{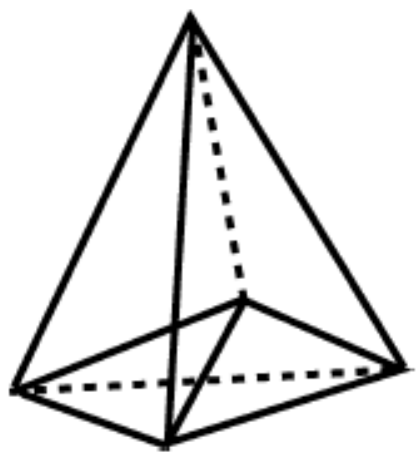}
\label{41simplex}
}
\caption[The four types of $4$-simplices employed in $(3+1)$-dimensional causal dynamical triangulations]{The four types of $4$-simplices employed in constructing $(3+1)$-dimensional causal triangulations: \subref{14simplex} $(1,4)$ $4$-simplex, \subref{23simplex} $(2,3)$ $4$-simplex, \subref{32simplex} $(3,2)$ $4$-simplex, \subref{41simplex} $(4,1)$ $4$-simplex. The first number in the ordered pair indicates the number of vertices on the initial spacelike hypersurface, and the second number in the ordered pair indicates the number of vertices on the final spacelike hypersurface. The future timelike direction points from the bottom to the top of the page. I have taken these images from \cite{JA&JJ&RL2}.}
\label{4simplices}
\end{figure}

The causal dynamical triangulations approach prescribes a rigorous definition of the path integration in equation \eqref{gravitypathintegral} that addresses all of these difficulties.\footnote{See \cite{JA&JJ&RL1,JA&JJ&RL2} for the original formulation and \cite{JA&AG&JJ&RL3} for a comprehensive review.}  One begins by asserting that the path integration in equation \eqref{gravitypathintegral} is restricted to the subclass of so-called causal spacetimes---those admitting a global foliation by spacelike hypersurfaces all of the same topology---for a definite choice of this topology. Accordingly, a causal spacetime manifold $\mathscr{M}_{c}$, equipped with a causal metric tensor $\mathbf{g}_{c}$, possesses the structure $\Sigma\times\mathcal{I}$, the direct product of a $d$-dimensional spatial manifold $\Sigma$ and a real temporal interval $\mathcal{I}$. One thus defines a quantum theory of gravity for each choice of $\Sigma$ 
within the causal dynamical triangulations approach. The relevant path integral for one such choice is then formally
\begin{equation}\label{causalgravitypathintegral}
\mathscr{A}_{\Sigma}[\gamma]=\int_{\substack{\mathscr{M}_{c}\cong\Sigma\times\mathcal{I} \\ \mathbf{g}_{c}|_{\partial\mathscr{M}_{c}}=\gamma}}\mathrm{d}\mu(\mathbf{g}_{c})\,e^{iS_{\mathrm{cl}}[\mathbf{g}_{c}]/\hbar}.
\end{equation}
One next invokes a lattice regularization of the path integration in equation \eqref{causalgravitypathintegral}. In particular, one replaces continuous causal spacetimes by so-called causal triangulations---piecewise-Minkowski simplicial manifolds possessing a distinguished global foliation by spacelike hypersurfaces all of the chosen topology. One constructs a causal triangulation $\mathcal{T}_{c}$ by appropriately gluing together the $d+1$ types of causal $(d+1)$-simplices, each a piece of Minkowski spacetime. In figure \ref{4simplices} I depict the four types of $4$-simplices. 
These $(d+1)$-simplices assemble so that each spacelike hypersurface is triangulated by regular spacelike $d$-simplices of squared edge length $\ell_{\mathrm{SL}}^{2}=a^{2}$, and adjacent spacelike hypersurfaces are connected by timelike edges of squared edge length $\ell_{\mathrm{TL}}^{2}=-\alpha a^{2}$ for real parameter $\alpha>0$. The \emph{a priori} arbitrary constant $a$ defines the lattice spacing. The connectivities of all $(d+1)$-simplices comprising a causal triangulation $\mathcal{T}_{c}$ together with these edge length assignments completely determine its spacetime geometry. This lattice regularization results in the replacement of the path integral \eqref{causalgravitypathintegral} by the path sum
\begin{equation}\label{causalgravitypathsum}
\mathcal{A}_{\Sigma}[\Gamma]=\sum_{\substack{\mathcal{T}_{c} \\ \mathcal{T}_{c}\cong\Sigma\times\mathcal{I} \\ \mathcal{T}_{c}|_{\partial\mathcal{T}_{c}}=\Gamma}}\mu(\mathcal{T}_{c})\,e^{i\mathcal{S}_{\mathrm{cl}}[\mathcal{T}_{c}]/\hbar}
\end{equation}
taken over all causal triangulations $\mathcal{T}_{c}$ isomorphic to $\Sigma\times\mathcal{I}$ with boundary $\Gamma$. The measure $\mu(\mathcal{T}_{c})$ is the inverse of the order of the automorphism group of the causal triangulation $\mathcal{T}_{c}$, and the action $\mathcal{S}_{\mathrm{cl}}[\mathcal{T}_{c}]$ is the translation of the action $S_{\mathrm{cl}}[\mathbf{g}_{c}]$ into the Regge calculus of causal triangulations. 

In most cases the path sum \eqref{causalgravitypathsum} is analytically intractable, so one studies the quantum theory so defined by numerical methods. In particular, one would like to perform Markov chain Monte Carlo simulations of representative paths contributing to the path sum \eqref{causalgravitypathsum}; however, such simulations cannot handle the complex amplitude $\mu(\mathcal{T}_{c})\,e^{i\mathcal{S}_{\mathrm{cl}}[\mathcal{T}_{c}]/\hbar}$ assigned to each path. To obtain a real amplitude for each path, one performs a Wick rotation of each causal triangulation from the Lorentzian to the Euclidean sector. This Wick rotation consists of the analytic continuation of $\alpha$ to $-\alpha$ through the lower half complex plane. The global foliability of each causal triangulation renders this Wick rotation globally well-defined. The Wick rotation transforms the path sum \eqref{causalgravitypathsum} into the partition function
\begin{equation}\label{partitionfunction}
\mathcal{Z}_{\Sigma}[\Gamma]=\sum_{\substack{\mathcal{T}_{c} \\ \mathcal{T}_{c}\cong\Sigma\times\mathcal{I} \\ \mathcal{T}_{c}|_{\partial\mathcal{T}_{c}}=\Gamma}}\mu(\mathcal{T}_{c})\,e^{-\mathcal{S}_{\mathrm{cl}}^{(\mathrm{E})}[\mathcal{T}_{c}]/\hbar},
\end{equation}
where $\mathcal{S}_{\mathrm{cl}}^{(\mathrm{E})}[\mathcal{T}_{c}]$ is the resulting Euclidean action. 
Since one can only run Markov chain Monte Carlo simulations of finite causal triangulations, one further specifies a fixed number $\bar{T}$ of spacelike hypersurfaces, introducing a discrete time coordinate $\tau$ to enumerate these time slices, and a fixed number $\bar{N}_{d+1}$ of $(d+1)$-simplices. 
One thus simulates representative paths contributing to the partition function
\begin{equation}\label{partitionfunctionfixedTN}
Z_{\Sigma}[\Gamma]=\sum_{\substack{\mathcal{T}_{c} \\ \mathcal{T}_{c}\cong\Sigma\times\mathcal{I} \\ \mathcal{T}_{c}|_{\partial\mathcal{T}_{c}}=\Gamma \\ T(\mathcal{T}_{c})=\bar{T} \\ N_{d+1}(\mathcal{T}_{c})=\bar{N}_{d+1}}}\mu(\mathcal{T}_{c})\,e^{-\mathcal{S}_{\mathrm{cl}}^{(\mathrm{E})}[\mathcal{T}_{c}]/\hbar},
\end{equation}
related by a Legendre transform to the partition function \eqref{partitionfunction}. 
A Markov chain Monte Carlo simulation outputs an ensemble $\mathcal{E}_{\Sigma}(\Gamma,\bar{T},\bar{N}_{4},\kappa_{0},\ldots,\kappa_{j},\alpha)$ of $N(\mathcal{T}_{c})$ causal triangulations 
for the chosen values of the bare couplings $\kappa_{0}$, $\ldots$ , $\kappa_{j}$ of the action $\mathcal{S}_{\mathrm{cl}}^{(\mathrm{E})}[\mathcal{T}_{c}]$. Appropriate powers of the lattice spacing render all of the bare couplings dimensionless. To estimate the expectation value of an observable $\mathcal{O}$ in the quantum state defined by the partition function \eqref{partitionfunctionfixedTN}, one computes its ensemble average
\begin{equation}\label{defensembleaverage}
\langle\mathcal{O}\rangle=\frac{1}{N(\mathcal{T}_{c})}\sum_{j=1}^{N(\mathcal{T}_{c})}\mathcal{O}_{j}.
\end{equation}
By studying the ensemble averages of sufficiently many observables, one characterizes completely this quantum state. 


One is ultimately interested in the hypothetical continuum limit of the quantum theory of gravity defined by the partition function \eqref{partitionfunctionfixedTN}. Typically, for a continuum limit to exist, the phase structure of the statistical model defined by the partition function \eqref{partitionfunctionfixedTN} must contain a second order phase transition. One expects to approach the continuum limit by letting the number $N_{d+1}$ of $(d+1)$-simplices increase without bound and the lattice spacing $a$ decrease to zero as one tunes the bare couplings to a fixed point of this second order phase transition. By studying the dependence of appropriate observables on the bare couplings, one explores the phase structure. One catalogues phases by identifying regions throughout each of which these observables exhibit a characteristic behavior, and one identifies transitions between phases by searching for changes in this characteristic behavior. Since one always works at finite $\bar{T}$ and $\bar{N}_{d+1}$, for which no true phase transitions occur, one employs finite size scaling techniques to extrapolate to the continuum limit. If a continuum limit exists, then the partition function \eqref{partitionfunctionfixedTN} defines a nonperturbatively renormalizable quantum theory of gravity in this limit. If a continuum limit does not exist, then the partition function \eqref{partitionfunctionfixedTN} defines an effective quantum theory of gravity valid on length scales greater than that of the regularization. One would like to study the renormalization group flows of either quantum theory of gravity. 



In the following I take 
\begin{equation}\label{EHaction}
S_{\mathrm{cl}}[\mathbf{g}]=\frac{1}{16\pi G_{0}}\int_{\mathscr{M}}\mathrm{d}^{4}x\,\sqrt{-g}\left(R-2\Lambda_{0}\right),
\end{equation}
the action of $4$-dimensional Einstein gravity in the absence of boundaries for bare Newton constant $G_{0}$ and bare cosmological constant $\Lambda_{0}$. 
For a causal spacetime manifold $\mathscr{M}_{c}$ of the form $\mathrm{S}^{3}\times\mathrm{S}^{1}$, the direct product of a spatial $3$-sphere and a temporal $1$-sphere, the action \eqref{EHaction} yields
\begin{eqnarray}\label{EReggeaction4}
\mathcal{S}_{\mathrm{cl}}^{(\mathrm{E})}[\mathcal{T}_{c}]&=&-(\kappa_{0}+6\Delta)N_{0}+\kappa_{4}\left(N_{4}^{(1,4)}+N_{4}^{(2,3)}+N_{4}^{(3,2)}+N_{4}^{(4,1)}\right)\nonumber\\ &&\qquad+\Delta\left(2N_{4}^{(1,4)}+N_{4}^{(2,3)}+N_{4}^{(3,2)}+2N_{4}^{(4,1)}\right)
\end{eqnarray}
in the Regge calculus of causal triangulations. The bare couplings $\kappa_{0}$, $\kappa_{4}$, and $\Delta$ are specific functions of $G_{0}/a^{2}$, $\Lambda_{0}a^{2}$, and $\alpha$; $N_{0}$ denotes the number of vertices; $N_{4}^{(p,q)}$ denotes the number of $(p,q)$ $4$-simplices. For this choice of spacetime topology, with the temporal interval $\mathcal{I}$ periodically identified, the partition function \eqref{partitionfunctionfixedTN} defines the ground state of the quantum geometry: there are no external sources in the form of a boundary condition $\mathcal{T}_{c}|_{\partial\mathcal{T}_{c}}=\Gamma$ perturbing the quantum geometry. 

\section{The phenomenology of causal dynamical triangulations}\label{phenomenology}

\subsection{Phases}\label{phases}

The particular quantum theory of gravity defined by the partition function \eqref{partitionfunctionfixedTN} for the action \eqref{EReggeaction4} exhibits three phases of quantum geometry: the decoupled phase A, the crumpled phase B, and the physical phase C \cite{JA&JJ&RL6,RK}. The A-C phase transition is of first order, the B-C phase transition is of second order, and the order of the A-B phase transition has not been ascertained \cite{JA&SJ&JJ&RL1,JA&SJ&JJ&RL2}. The three phases are hypothesized to meet at a triple point \cite{CDTandHL}. My proposed renormalization group scheme applies only to phase C. 
For this phase there are several observables whose extensive study has linked their phenomenology to a theoretical model, allowing for the formulation of a renormalization group scheme. 
I thus concentrate on phase C in the remainder of section \ref{phenomenology} and in section \ref{renormalization}, returning to considerations of the phase transitions in section \ref{conclusion}. 

\subsection{Observables}\label{observables}

Of these several observables for phase C, I focus on three in particular: the $1$-point function of the discrete spatial volume as a function of the discrete time coordinate, the connected $2$-point function of fluctuations in the discrete spatial volume as a function of the discrete time coordinate, and the spectral dimension as a function of diffusion time. These three observables were studied first in \cite{JA&JJ&RL1,JA&AG&JJ&RL1,JA&JJ&RL7} (respectively) and subsequently in \cite{JA&AG&JJ&RL2,JA&AG&JJ&RL&JGS&TT,JA&JJ&RL5,JA&JJ&RL6,CA&SJC&JHC&PH&RKK&PZ,DB&JH,JHC&JMM,RK}. I successively discuss the phenomenology of these three observables in subsubsections \ref{1ptfunction}, \ref{2ptfunction}, and \ref{SpectralDimension}, paying particular attention to the length scales that each sets. The emphasis on dynamically emergent scales is the primary novel feature of this otherwise standard account. Unless noted by a citation, the numerical data newly presented here and in subsection \ref{model} was generated from Markov chain Monte Carlo simulations using the code reported in \cite{RK}. 

\subsubsection{$1$-point function of the discrete spatial volume}\label{1ptfunction}

\begin{figure}[!ht]
\centering
\subfigure[ ]{
\includegraphics[scale=0.6]{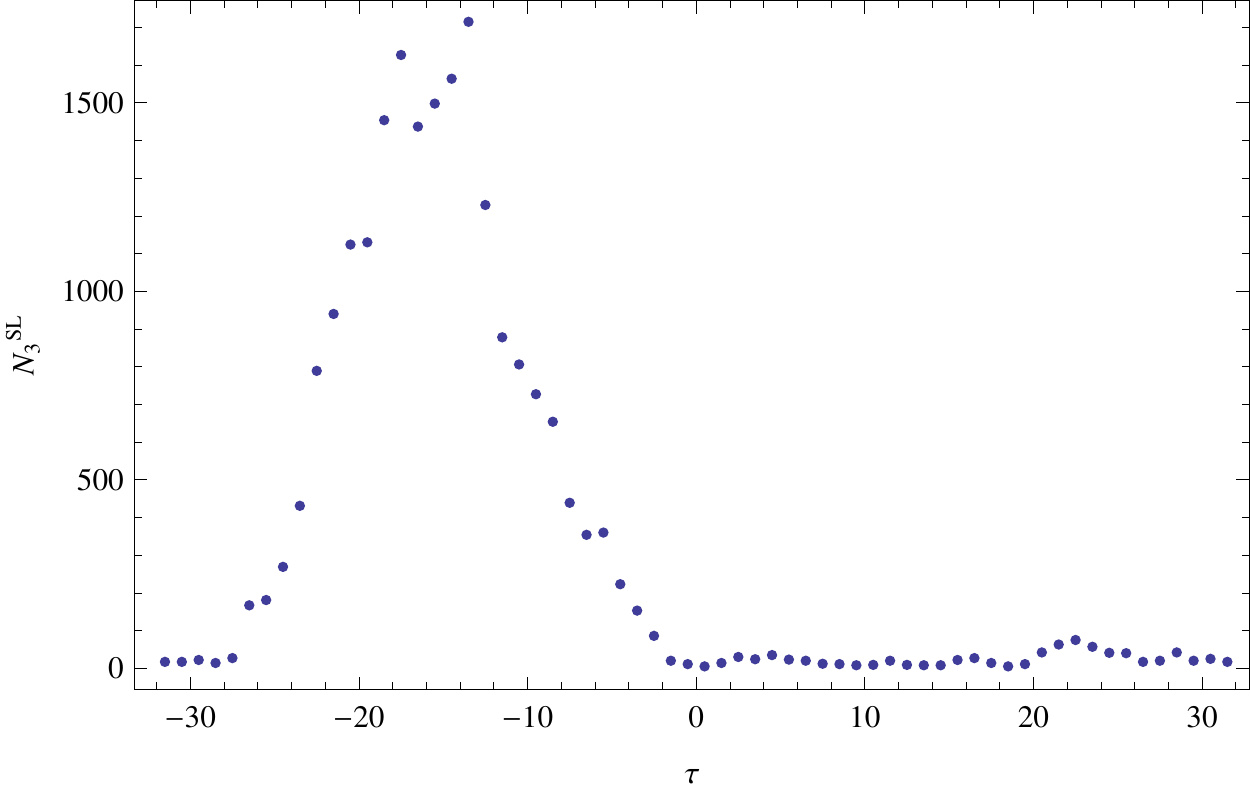}
\label{singlevolprof3p1T64V80k}
}
\subfigure[ ]{
\includegraphics[scale=0.6]{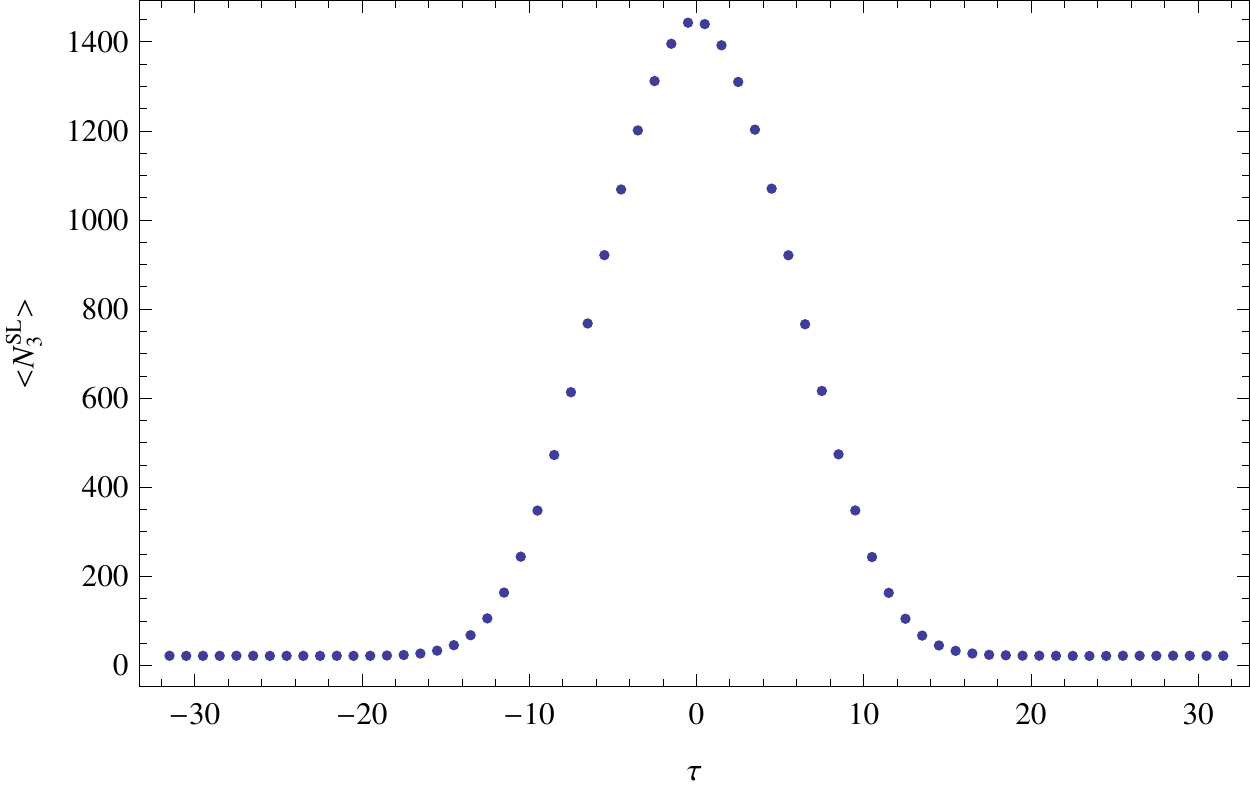}
\label{volprof3p1T64V80k}
}
\caption{\subref{singlevolprof3p1T64V80k} The number $N_{3}^{\mathrm{SL}}$ of spacelike $3$-simplices as a function of the discrete time coordinate $\tau$ for a representative causal triangulation from an ensemble within phase C characterized by $\bar{T}=64$, $\bar{N}_{4}=81920$, $\kappa_{0}=2.0$, and $\Delta=0.4$. \subref{volprof3p1T64V80k} The coherent ensemble average number $\langle N_{3}^{\mathrm{SL}}\rangle$ of spacelike $3$-simplices as a function of the discrete time coordinate $\tau$ for an ensemble within phase C characterized by $\bar{T}=64$, $\bar{N}_{4}=81920$, $\kappa_{0}=2.0$, and $\Delta=0.4$.}
\label{volprofiles}
\end{figure}

I measure the discrete spatial $3$-volume as the number $N_{3}^{\mathrm{SL}}(\tau)$ of spacelike $3$-simplices comprising the time slice labeled by the discrete time coordinate $\tau$. One may conceive of $N_{3}^{\mathrm{SL}}(\tau)$ as the temporal evolution of the discrete spatial $3$-volume. Following the prescription \eqref{defensembleaverage}, I estimate the $1$-point function by the ensemble average $\langle N_{3}^{\mathrm{SL}}(\tau)\rangle$. For a spacetime manifold of the form $\mathrm{S}^{3}\times\mathrm{S}^{1}$, one performs this ensemble average coherently as explained, for instance, in \cite{CA&SJC&JHC&PH&RKK&PZ}. 
In figure \ref{volprofiles}\subref{singlevolprof3p1T64V80k} I display $N_{3}^{\mathrm{SL}}(\tau)$ for a representative causal triangulation from a typical ensemble within phase C. In figure \ref{volprofiles}\subref{volprof3p1T64V80k} I display $\langle N_{3}^{\mathrm{SL}}(\tau)\rangle$ for this same ensemble. 
Clearly, $N_{3}^{\mathrm{SL}}(\tau)$ possesses a distinct form within phase C. There is a single accumulation of discrete spatial $3$-volume spanning a significant portion of the time slices. In figure \ref{volprofiles}\subref{singlevolprof3p1T64V80k}, for instance, this accumulation spans the time slices for which $\tau\in\{-26,\ldots,-3\}$. Within this accumulation the discrete spatial $3$-volume exhibits a temporal modulation from small to large to small values. Since I center this accumulation in the coherent ensemble average, as in figure \ref{volprofiles}\subref{volprof3p1T64V80k}, I refer to it as the central accumulation. The time slices beyond the central accumulation constitute the so-called stalk in which each time slice has a near minimal number of spacelike $3$-simplices. In figure \ref{volprofiles}\subref{singlevolprof3p1T64V80k}, for instance, the stalk spans the time slices for which $\tau\in\{-32,\ldots,-27\}\cup\{-2,\ldots,32\}$. The stalk is a numerical artifact stemming from the requirement that every time slice possesses the topology of a $3$-sphere \cite{JHC&JMM}. Accordingly, the central accumulation alone is considered to constitute the physical portion of a causal triangulation in phase C. 

The central accumulation appears to set two scales, its extent and its duration. As expected, based on numerical measurements alone, both of these scales are characterized by dimensionless numbers: the extent is just a certain number of spacelike $3$-simplices, and the duration is just a certain number of time slices. One can attempt to express the extent and the duration in units of the lattice spacing $a$, but the lattice spacing is an \emph{a priori} arbitrary scale, as I emphasized previously. In section \ref{model} I bring the extent and the duration into contact with a particular model, which shows these two scales to be in fact just one. This one scale, which I presciently denote $\ell_{\mathrm{dS}}$, is the largest scale characterizing the ensemble average quantum geometry in phase C. 

\subsubsection{$2$-point function of fluctuations in the discrete spatial volume}\label{2ptfunction}

\begin{figure}[!ht]
\centering
\subfigure[ ]{
\includegraphics[scale=0.6]{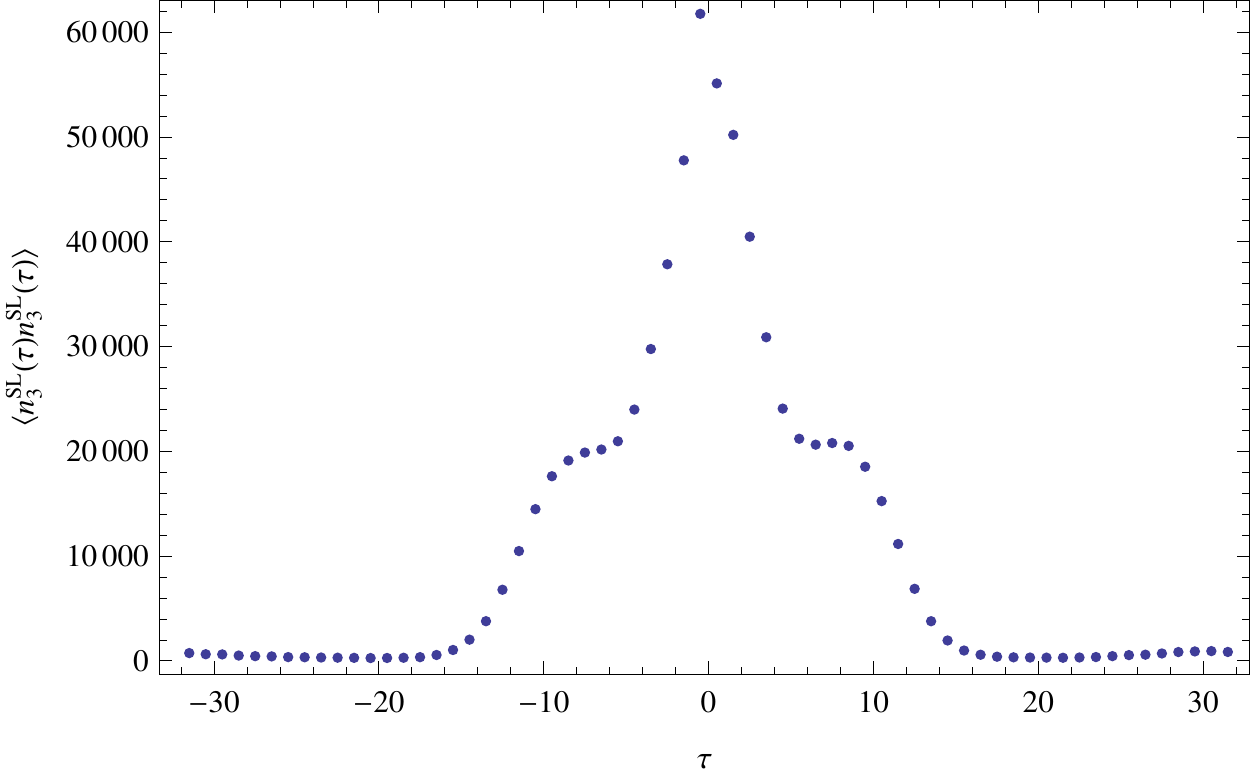}
\label{covariancediagonal}
}
\subfigure[ ]{
\includegraphics[scale=0.6]{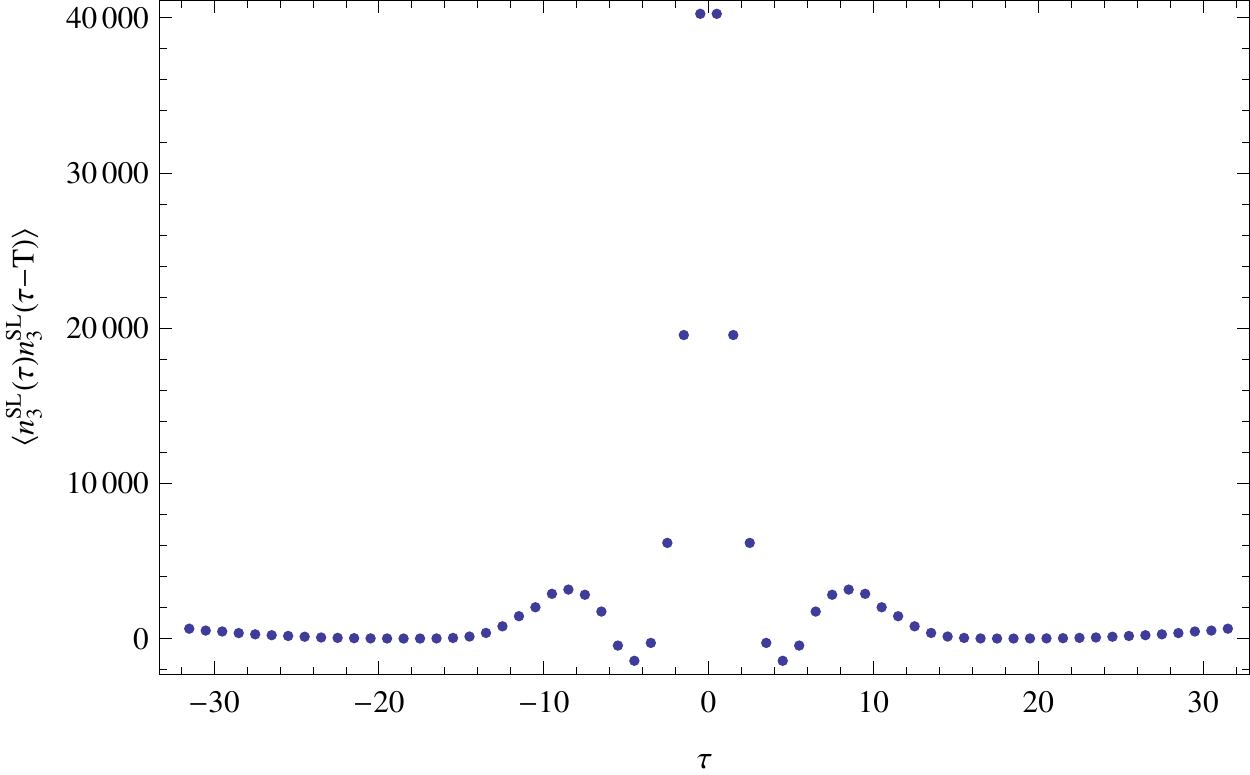}
\label{covarianceantidiagonal}
}
\caption{The connected $2$-point function $\langle n_{3}^{\mathrm{SL}}(\tau)n_{3}^{\mathrm{SL}}(\tau')\rangle$ of fluctuations $n_{3}^{\mathrm{SL}}(\tau)$ in the number $N_{3}^{\mathrm{SL}}(\tau)$ of spacelike $3$-simplices for an ensemble within phase C characterized by $\bar{T}=64$, $\bar{N}_{4}=81920$, $\kappa_{0}=2.0$, and $\Delta=0.4$. \subref{covariancediagonal} The diagonal $\langle n_{3}^{\mathrm{SL}}(\tau)n_{3}^{\mathrm{SL}}(\tau)\rangle$. \subref{covarianceantidiagonal} The antidiagonal $\langle n_{3}^{\mathrm{SL}}(\tau)n_{3}^{\mathrm{SL}}(\tau-T)\rangle$. }
\label{2ptfunction3p1T64V80}
\end{figure}

I measure the fluctuation in the discrete spatial $3$-volume as a function of the discrete time coordinate as
\begin{equation}
\left[n_{3}^{\mathrm{SL}}(\tau)\right]_{j}=\left[N_{3}^{\mathrm{SL}}(\tau)\right]_{j}-\langle N_{3}^{\mathrm{SL}}(\tau)\rangle
\end{equation}
for member $j$ of an ensemble of $N(\mathcal{T}_{c})$ causal triangulations. Again following the prescription \eqref{defensembleaverage}, I estimate the connected $2$-point function of these fluctuations by the ensemble average
\begin{equation}\label{def2ptfunction}
\langle n_{3}^{\mathrm{SL}}(\tau)n_{3}^{\mathrm{SL}}(\tau')\rangle=\frac{1}{N(\mathcal{T}_{c})}\sum_{j=1}^{N(\mathcal{T}_{c})}\left[n_{3}^{\mathrm{SL}}(\tau)\right]_{j}\left[n_{3}^{\mathrm{SL}}(\tau')\right]_{j}.
\end{equation}
For a spacetime manifold of the form $\mathrm{S}^{3}\times\mathrm{S}^{1}$, one respects the coherent ensemble averaging of $N_{3}^{\mathrm{SL}}(\tau)$ when computing the $2$-point function \eqref{def2ptfunction}. 
One may conceive of $\langle n_{3}^{\mathrm{SL}}(\tau)n_{3}^{\mathrm{SL}}(\tau')\rangle$ as the temporal correlator or propagator of fluctuations in the discrete spatial $3$-volume. For a typical ensemble of causal triangulations within phase C, I display the diagonal $\langle n_{3}^{\mathrm{SL}}(\tau)n_{3}^{\mathrm{SL}}(\tau)\rangle$ in figure \ref{2ptfunction3p1T64V80}\subref{covariancediagonal} and the antidiagonal $\langle n_{3}^{\mathrm{SL}}(\tau)n_{3}^{\mathrm{SL}}(\tau-T)\rangle$ in figure \ref{2ptfunction3p1T64V80}\subref{covarianceantidiagonal}. 

\begin{figure}[!ht]
\centering
\includegraphics[width=\linewidth]{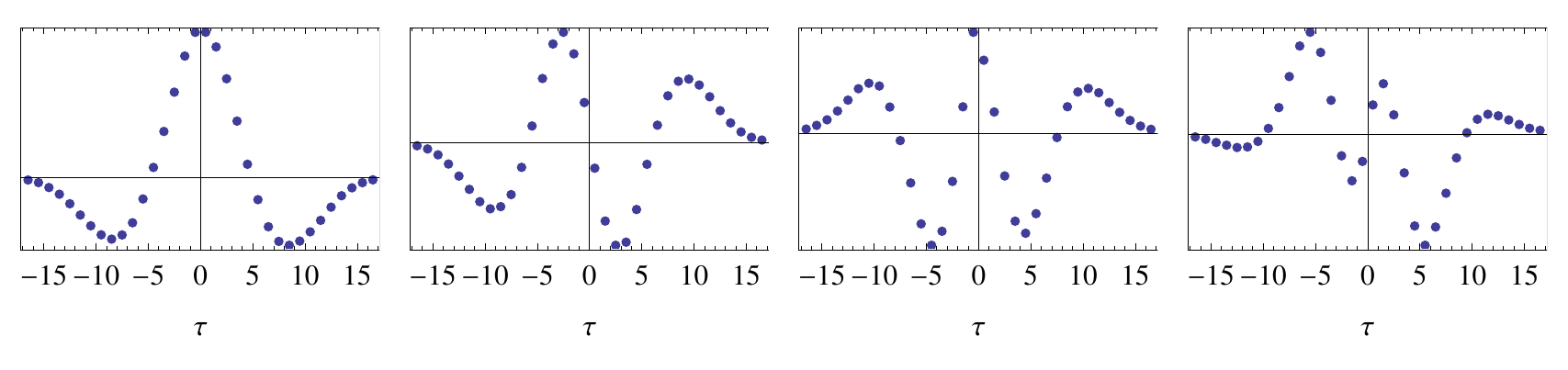}
\caption{The first four eigenvectors $[\eta_{3}^{\mathrm{SL}}(\tau)]_{j}$ of the connected $2$-point function $\langle n_{3}^{\mathrm{SL}}(\tau)n_{3}^{\mathrm{SL}}(\tau')\rangle$ within the central accumulation for an ensemble within phase C characterized by $\bar{T}=64$, $\bar{N}_{4}=81920$, $\kappa_{0}=2.0$, and $\Delta=0.4$.}
\label{eigenvectors}
\end{figure}

\begin{figure}[!ht]
\centering
\includegraphics[scale=0.6]{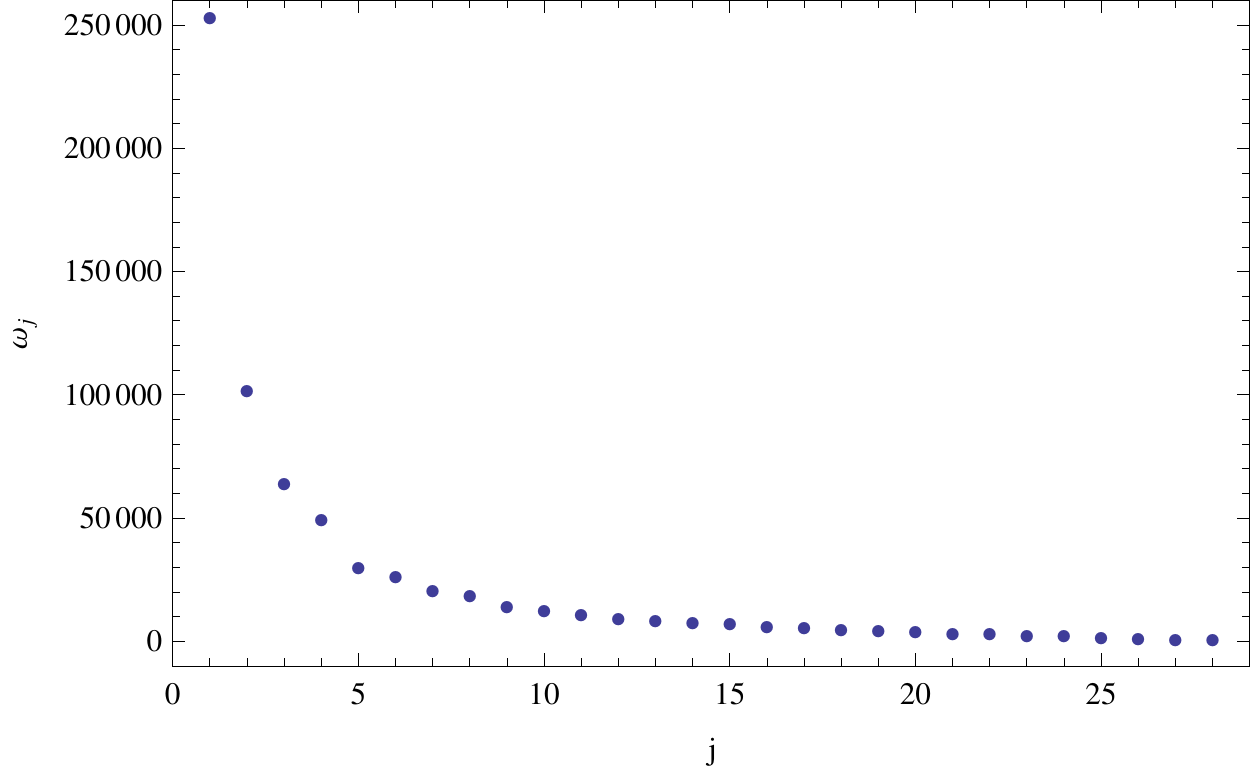}
\caption{The eigenvalues $\omega_{j}$ of the connected $2$-point function $\langle n_{3}^{\mathrm{SL}}(\tau)n_{3}^{\mathrm{SL}}(\tau')\rangle$ within the central accumulation for an ensemble within phase C characterized by $\bar{T}=64$, $\bar{N}_{4}=81920$, $\kappa_{0}=2.0$, and $\Delta=0.4$.}
\label{eigenvalues}
\end{figure}

\begin{figure}[!ht]
\centering
\includegraphics[width=\linewidth]{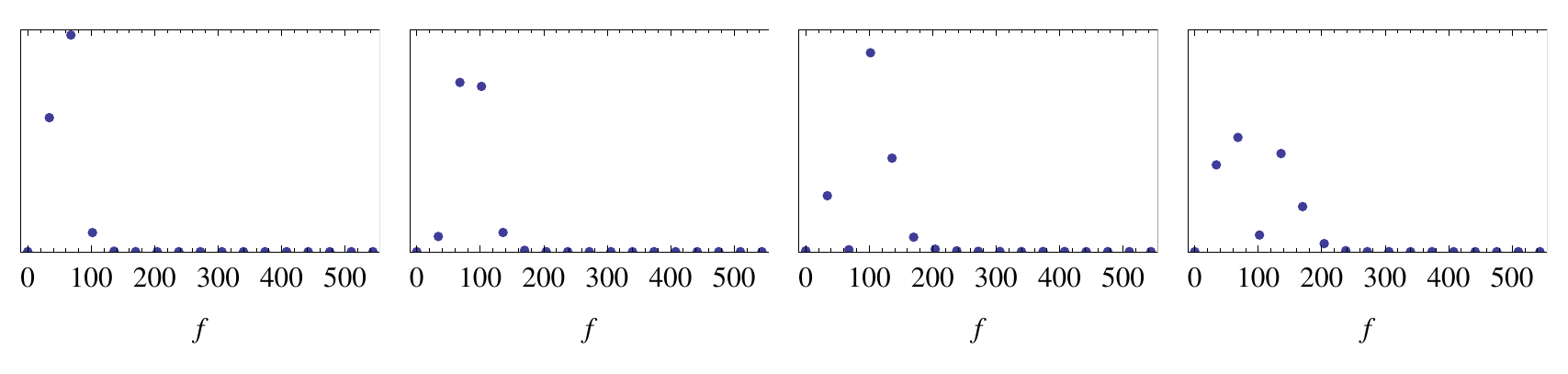}
\caption{The power spectra computed by discrete Fourier transform of the first four eigenvectors $[\eta_{3}^{\mathrm{SL}}(\tau)]_{j}$ of the connected $2$-point function $\langle n_{3}^{\mathrm{SL}}(\tau)n_{3}^{\mathrm{SL}}(\tau')\rangle$ within the central accumulation for an ensemble within phase C characterized by $\bar{T}=64$, $\bar{N}_{4}=81920$, $\kappa_{0}=2.0$, and $\Delta=0.4$.}
\label{eigenvectorspectra}
\end{figure}

One may decompose $\langle n_{3}^{\mathrm{SL}}(\tau)n_{3}^{\mathrm{SL}}(\tau')\rangle$---just a symmetric $T\times T$ matrix---into a spectrum of $T$ eigenvectors $\left[\eta_{3}^{\mathrm{SL}}(\tau)\right]_{j}$ with associated eigenvalues $\omega_{j}$. In figure \ref{eigenvectors} I display the first four eigenvectors $\left[\eta_{3}^{\mathrm{SL}}(\tau)\right]_{j}$, and in figure \ref{eigenvalues} I display the spectrum of eigenvalues $\omega_{j}$. 
I have restricted consideration to the central accumulation so that $\langle n_{3}^{\mathrm{SL}}(\tau)n_{3}^{\mathrm{SL}}(\tau')\rangle$ is a $T_{\mathrm{ph}}\times T_{\mathrm{ph}}$ matrix. The central accumulation spans $T_{\mathrm{ph}}$ time slices. As I demonstrate in section \ref{model}, the eigenvectors $\left[\eta_{3}^{\mathrm{SL}}(\tau)\right]_{j}$ constitute a finite set of excitations of the quantum geometry, 
and the inverses of the eigenvalues $\omega_{j}$ measure the effective spacetime energy-momentum density of these excitations. Decomposing each eigenvector $\left[\eta_{3}^{\mathrm{SL}}(\tau)\right]_{j}$ into its discrete Fourier modes, one obtains its dominant frequency $f_{j}$ or period $p_{j}$, again characterized by a dimensionless number. In figure \ref{eigenvectorspectra} I display the power spectra of the discrete Fourier transforms of the first four eigenvectors $\left[\eta_{3}^{\mathrm{SL}}(\tau)\right]_{j}$. 
Numerical measurements are sensitive to periods neither less than twice the spacing $\Delta\tau$ between adjacent time slices nor greater than half the number $T_{\mathrm{ph}}$ of time slices. The dominant periods $p_{j}$ of the eigenvectors $\left[\eta_{3}^{\mathrm{SL}}(\tau)\right]_{j}$ thus define a ladder of scales $\ell_{p_{j}}$, each a fraction of the extent $\ell_{\mathrm{dS}}$, descending from the largest scale $\ell_{p_{1}}$ to the successively shorter scales $\ell_{p_{2}},\ldots,\ell_{p_{T_{\mathrm{ph}}}}$. The eigenvalues $\omega_{j}$ also define a ladder of scales $\ell_{\omega_{j}}$---the seventh roots of the eigenvalues $\omega_{j}$---extending over an order of magnitude. Again, as expected, based on numerical measurements alone, all of these scales are characterized by dimensionless numbers. In subsection \eqref{model} I bring the scales $\ell_{p_{j}}$ and $\ell_{\omega_{j}}$ into contact with a particular model, which allows for their expression in units of the lattice spacing $a$.

\subsubsection{Spectral dimension}\label{SpectralDimension}

The spectral dimension $d_{\mathfrak{s}}(\sigma)$ of a space measures its effective dimensionality on the scale $\sigma$ as witnessed by a random walker that has diffused for a time $\sigma$. If the space is a causal triangulation, then the integrated heat equation describing this diffusion process takes the form
\begin{equation}\label{heatequation}
K(s,s',\sigma+1)=(1-\varrho)K(s,s',\sigma)+\frac{\varrho}{N(\mathscr{N}(s))}\sum_{s''\in\mathscr{N}(s)}K(s'',s',\sigma)
\end{equation}
subject to the initial condition
\begin{equation}\label{heatequationinitialcondition}
K(s,s',0)=\delta_{ss'}.
\end{equation}
The heat kernel $K(s,s',\sigma)$ gives the probability for the random walker to diffuse from $4$-simplex $s$ to $4$-simplex $s'$ (or \emph{vice versa}) in $\sigma$ diffusion time steps. The diffusion constant $\varrho$ characterizes the dwell probability in any given step of the diffusion process. $\mathscr{N}(s)$ is the set of $N(\mathscr{N}(s))$ $4$-simplices that neighbor the $4$-simplex $s$. For a spacetime manifold of the form $\mathrm{S}^{3}\times\mathrm{S}^{1}$, $N(\mathscr{N}(s))=5$ for all $4$-simplices in a causal triangulation. From the heat kernel $K(s,s',\sigma)$ one computes the heat trace or return probability as
\begin{equation}
P(\sigma)=\frac{1}{N_{4}}\sum_{s\in\mathcal{T}_{c}}K(s,s,\sigma),
\end{equation}
which gives the probability for the random walker to return to its starting $4$-simplex in $\sigma$ diffusion time steps. One then defines the spectral dimension as
\begin{equation}\label{specdim}
d_{\mathfrak{s}}(\sigma)=-2\frac{\mathrm{d}\ln{P(\sigma)}}{\mathrm{d}\ln{\sigma}}
\end{equation}
for an appropriate discretization of the derivative with respect to diffusion time. 

\begin{figure}[!ht]
\centering
\subfigure[ ]{
\includegraphics[scale=0.525]{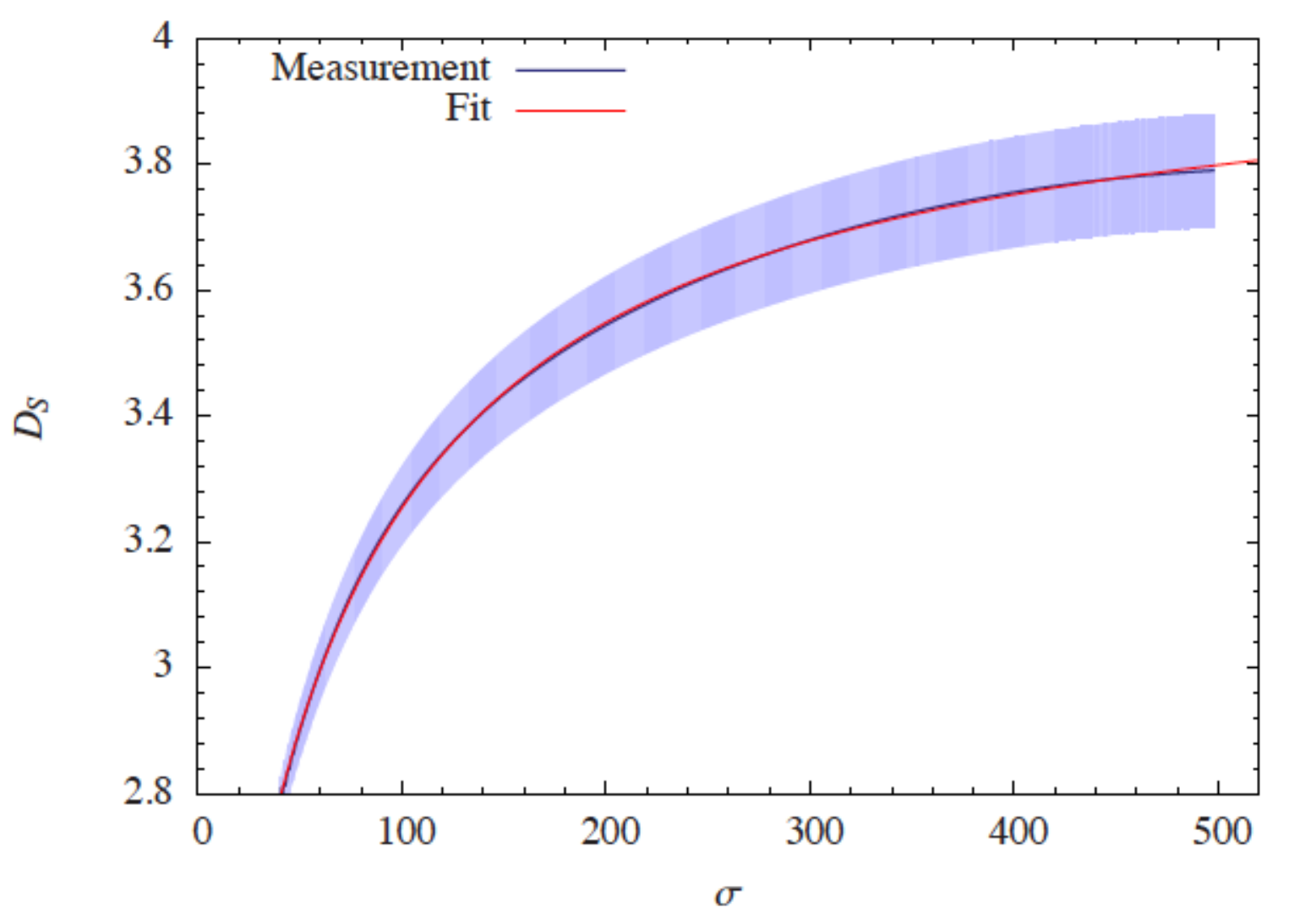}
\label{AGJLspecdim}
}
\subfigure[ ]{
\includegraphics[scale=0.7]{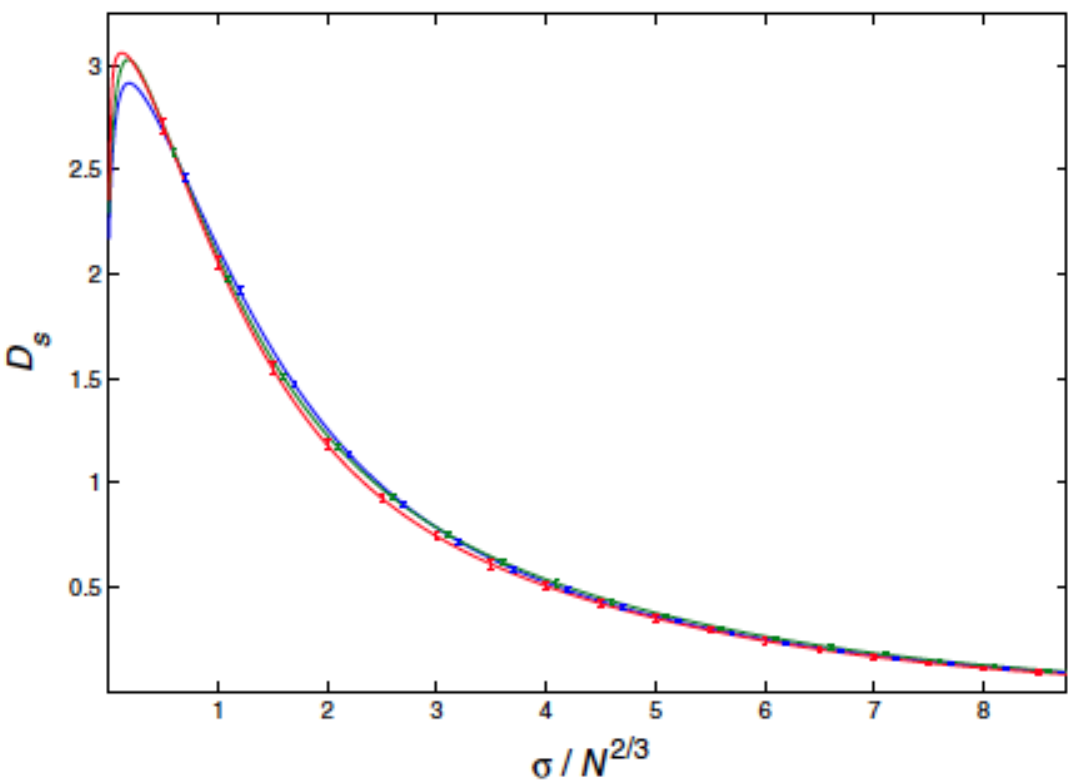}
\label{BHspecdim}
}
\caption{\subref{AGJLspecdim} The ensemble average spectral dimension $\langle d_{\mathfrak{s}}\rangle$ as a function of diffusion time $\sigma$ for an ensemble of $(3+1)$-dimensional causal triangulations within phase C. I have taken this image from \cite{JA&AG&JJ&RL3}. \subref{BHspecdim} The ensemble average spectral dimension $\langle d_{\mathfrak{s}}\rangle$ as a function of finite size scaled diffusion time $\sigma/N_{3}^{2/3}$ for an ensemble of $(2+1)$-dimensional causal triangulations within phase C. I have taken this image from \cite{DB&JH}.}
\label{spectraldimension}
\end{figure}

Once again following the prescription \eqref{defensembleaverage}, I estimate the expectation value of the return probability by the ensemble average
\begin{equation}
\langle P(\sigma)\rangle=\frac{1}{N(\mathcal{T}_{c})}\sum_{j=1}^{N(\mathcal{T}_{c})}P_{j}(\sigma).
\end{equation}
I measure the return probability $P_{j}(\sigma)$ by iteration of equation \eqref{heatequation} starting from the initial condition \eqref{heatequationinitialcondition}. I then compute the ensemble average spectral dimension $\langle d_{\mathfrak{s}}(\sigma)\rangle$ according to the definition \eqref{specdim}. In figure \ref{spectraldimension} I display $\langle d_{\mathfrak{s}}(\sigma)\rangle$ for a typical ensemble of causal triangulations within phase C. 
Clearly, $\langle d_{\mathfrak{s}}(\sigma)\rangle$ exhibits a distinct form within phase C. Let $\sigma_{\mathrm{cl}}$ denote the diffusion time at which $\langle d_{\mathfrak{s}}(\sigma)\rangle$ attains its maximum. For instance, in figure \ref{spectraldimension}\subref{AGJLspecdim} $\sigma_{\mathrm{cl}}$ slightly exceeds $500$, and in figure \ref{spectraldimension}\subref{BHspecdim} $\sigma_{\mathrm{cl}}$ is approximately $300$ (corresponding to a finite size scale diffusion time of approximately $5\cdot10^{-3}$).
For $\sigma\in(0,\sigma_{\mathrm{cl}})$ the spectral dimension increases monotonically from approximately $2$ to approximately $4$ as depicted in figure \ref{spectraldimension}\subref{AGJLspecdim}. This is the regime in which quantum-mechanical effects dominate the diffusion process, giving rise to the phenomenon of dynamical dimensional reduction \cite{JA&JJ&RL7}. In the vicinity of $\sigma_{\mathrm{cl}}$, the spectral dimension is approximately constant at  the topological value of $4$. This is presumably the semiclassical regime in which the ensemble average quantum geometry possibly appears approximately Euclidean to the random walker. For $\sigma\in(\sigma_{\mathrm{cl}},\infty)$ the spectral dimension gradually decays to zero as depicted in figure \ref{spectraldimension}\subref{BHspecdim}. This is the regime in which large scale positive curvature dominates the diffusion process. The diffusion time $\sigma_{\mathrm{cl}}$ sets the scale $\ell_{\mathrm{cl}}$ on which the quantum geometry becomes approximately classical. As I demonstrate in section \ref{model}, the rate of decay for $\sigma>\sigma_{\mathrm{cl}}$ sets the scale of the large scale curvature, which is also the scale $\ell_{\mathrm{dS}}$ that characterizes the central accumulation.\footnote{Numerical measurements of $\langle d_{\mathfrak{s}}(\sigma)\rangle$ interpreted within the model of subsection \ref{model} do not alone determine the scale $\ell_{\mathrm{dS}}$ in units of the lattice spacing $a$.} If one possessed a model of the spectral dimension's behavior for $\sigma<\sigma_{\mathrm{cl}}$, of which there are several \cite{JA&JJ&RL7,GC1,GC2,GC&AE&FS,GC&GN,SC2,PetrSpec,FracSpace,SR&FS,MR&FS,Visser3}, then one could presumably also extract a scale $\ell_{\mathrm{qm}}$ from the rate of increase. Once again, as expected, based on numerical measurements alone, these three scales are characterized by dimensionless numbers. 

Unlike the scales $\ell_{\mathrm{dS}}$, $\ell_{p_{j}}$, and $\ell_{\omega_{j}}$, which can be determined in units of the lattice spacing $a$ \emph{via} the model introduced in subsection \ref{model}, the scales $\ell_{\mathrm{cl}}$ and $\ell_{\mathrm{qm}}$ are not easily connected to the lattice spacing $a$. The diffusion time $\sigma$ simply counts the number of steps that the random walker has taken, and this number is not tied \emph{a priori} to the lattice spacing. In a forthcoming paper I propose a method for determining the equivalent of $\sigma$ diffusion time steps in units of the lattice spacing $a$ \cite{JHC2}. Accordingly, in the following I assume that one can determine $\ell_{\mathrm{cl}}$ and $\ell_{\mathrm{qm}}$ in terms of the lattice spacing $a$. My renormalization group scheme does not necessarily depend on this assumption.

\subsection{Model}\label{model}

I now attempt to account for the phenomenology of the observables discussed in subsection \ref{observables} with a simple model. The model presented below was first formulated in \cite{JA&JJ&RL5} and further studied in \cite{JA&JGS&AG&JJ,JA&AG&JJ&RL1,JA&AG&JJ&RL2,JA&AG&JJ&RL&JGS&TT,JA&JJ&RL6}. 

By measuring the discrete spatial $3$-volume $N_{3}^{\mathrm{SL}}(\tau)$ as a function of the discrete time coordinate $\tau$, I ignore all other geometrical information characterizing the time slices, and I probe the quantum geometry only on its largest scales. Accordingly, I should consider a model in which the metric tensor $\mathbf{g}$ is characterized solely by the spatial $3$-volume as a function of coordinate time---a minisuperspace approximation of the metric degrees of freedom---and in which the interactions of the metric tensor $\mathbf{g}$ are those most relevant on large scales---the Einstein-Hilbert truncation of the gravitational effective action. Since I study causal dynamical triangulations in its Euclidean sector, I take as my model the minisuperspace approximation of the Euclidean Einstein-Hilbert truncation. 
The metric tensor  $\mathbf{g}$ is thus specified by the line element
\begin{equation}\label{lineelement}
\mathrm{ds}^{2}=g_{tt}\mathrm{d}t^{2}+\rho^{2}(t)\left[\mathrm{d}\psi^{2}+\sin^{2}{\psi}\left(\mathrm{d}\theta^{2}+\sin^{2}{\theta}\,\mathrm{d}\varphi^{2}\right)\right]
\end{equation}
for real constant $g_{tt}$ and scale factor $\rho(t)$. In terms of the spatial $3$-volume
\begin{equation}
V_{3}(t)=\int_{0}^{\pi}\mathrm{d}\psi\int_{0}^{\pi}\mathrm{d}\theta\int_{0}^{2\pi}\mathrm{d}\varphi\,\rho^{3}(t)\sin^{2}{\psi}\sin{\theta}=2\pi^{2}\rho^{3}(t),
\end{equation}
the action \eqref{EHaction}, Wick rotated to the Euclidean sector, becomes
\begin{equation}\label{MSM3action4}
S^{(\mathrm{E})}[V_{3}]=\frac{1}{24\pi G_{\mathrm{ren}}}\int\mathrm{d}t\sqrt{g_{tt}}\left(\frac{\dot{V}_{3}^{2}}{g_{tt}V_{3}}+2^{2/3}3^{2}\pi^{4/3}V_{3}^{1/3}-3\tilde{\Lambda} V_{3}\right)
\end{equation}
after integrating by parts.\footnote{Technically, a standard Wick rotation of the action \eqref{EHaction} yields an overall negative sign, giving a kinetic term of the wrong sign. This is the well-known unboundedness from below of the conformal mode of the metric tensor $\mathbf{g}$. Evidently, the effective action describing the ensemble average quantum geometry of phase C on sufficiently large scales does not have this overall negative sign \cite{JA&JJ&RL5,JA&JJ&RL6}.} 
I work with the spatial $3$-volume $V_{3}(t)$, not the scale factor $\rho(t)$, because the former, not the latter, is the analogue of $N_{3}^{\mathrm{SL}}(\tau)$ in the naive continuum limit. Since one runs the Markov chain Monte Carlo simulations described in section \ref{theory} at fixed number $\bar{N}_{4}$ of $4$-simplices, I add the term 
\begin{equation}\label{MSM3action4c}
S_{V_{4}}^{(\mathrm{E})}[V_{3}]=\lambda\left[\int\mathrm{d}t\sqrt{g_{tt}}V_{3}-\bar{V}_{4}\right]
\end{equation}
to the action \eqref{MSM3action4} to constrain the spacetime $4$-volume $V_{4}$ to equal $\bar{V}_{4}$. The term \eqref{MSM3action4c} simply leads to an effective cosmological constant $\Lambda_{\mathrm{ren}}=\tilde{\Lambda}-8\pi G_{\mathrm{ren}}\lambda$ shifted by the Lagrange multiplier $\lambda$. 

The effective action \eqref{MSM3action4}, as opposed to the bare action \eqref{EHaction}, is intended to describe the dynamics emerging from the partition function \eqref{partitionfunctionfixedTN} on sufficiently large scales. Its couplings are the renormalized Newton constant $G_{\mathrm{ren}}$ and the renormalized cosmological constant $\Lambda_{\mathrm{ren}}$ on these scales. In calling $G_{\mathrm{ren}}$ the Newton constant, I tacitly assume that this coupling in the action \eqref{MSM3action4} functions in the capacity of the Newton constant. Besides the expectation that the Newton constant sets the scale of fluctuations in the spatial $3$-volume, which I demonstrate in subsubsection \ref{modelobservables}, this assumption is based exclusively on identification of symbols. Ideally, one should attempt to verify that $G_{\mathrm{ren}}$ characterizes, for instance, the strength of the gravitational interaction between two massive bodies within phase C. Such a demonstration, however, is highly nontrivial. In calling $\Lambda_{\mathrm{ren}}$ the cosmological constant, I also tacitly assume that this coupling in the action \eqref{MSM3action4} functions in the capacity of the cosmological constant. This assumption appears to be well-justified based on the interpretation in subsubsection \ref{modelobservables} of the observables discussed in subsection \ref{observables}.

That a minisuperspace approximation of the Einstein-Hilbert truncation provides a good model of my numerical measurements on sufficiently large scales is notable yet expected: causal dynamical triangulations quite possibly possesses the correct classical limit, but of course the two most relevant terms in the gravitational effective action 
should dominate on these scales. 


\subsubsection{Finite size scaling towards the continuum limit}

To model the discrete observables of subsection \ref{observables} on the basis of the continuous action \eqref{MSM3action4}, I require some means to connect the discrete with the continuous. A finite size scaling \emph{Ansatz} dictates the transformation of discrete quantities into their continuous counterparts in the continuum limit. In the setting of causal dynamical triangulations, one naively expects to approach the continuum limit by letting the number $N_{4}$ increase without bound and the lattice spacing $a$ decrease towards zero while the product $N_{4}a^{4}$ remains constant. This expectation leads to the canonical finite size scaling \emph{Ansatz}: in the continuum limit the spacetime $4$-volume $V_{4}$ is given by
\begin{equation}\label{naiveCL}
V_{4}=\lim_{\substack{N_{4}\rightarrow\infty \\ a\rightarrow 0}}C_{4}N_{4}a^{4}.
\end{equation}
The constant $C_{4}$ is the discrete spacetime $4$-volume of an effective $4$-simplex in an ensemble of causal triangulations: 
\begin{equation}\label{C4}
C_{4}=\frac{1}{\bar{N}_{4}a^{4}}\left(\langle N_{4}^{(1,4)}\rangle V_{4}^{(1,4)}+\langle N_{4}^{(2,3)}\rangle V_{4}^{(2,3)}+\langle N_{4}^{(3,2)}\rangle V_{4}^{(3,2)}+\langle N_{4}^{(4,1)}\rangle V_{4}^{(4,1)}\right)
\end{equation}
for the Euclidean spacetime $4$-volume $V_{4}^{(p,q)}$ of a $(p,q)$ $4$-simplex. Since the product $C_{4}N_{4}a^{4}$ remains fixed as one approaches the continuum limit, one expects the relation \eqref{naiveCL} to hold also for finite $N_{4}$ and $a$. Typically, the relation \eqref{naiveCL} only obtains for finite $N_{4}$ and $a$ within an appropriate scaling regime, which fortunately appears to constitute a large portion of phase C. In practice, one expects subleading corrections to the \emph{Ansatz} \eqref{naiveCL} for finite $N_{4}$ and $a$ of the form
\begin{equation}
V_{4}=C_{4}N_{4}a^{4}+O(N_{4}^{1-\epsilon},a^{4+\epsilon}).
\end{equation}
These subleading corrections reinforce that the \emph{Ansatz} \eqref{naiveCL} is a hypothesis that one should test against statistical analyses of numerical measurements. Near the boundaries of phase C, particularly that with phase B, the finite size scaling \emph{Ansatz} might receive substantial subleading or even leading corrections. Indeed, if the second order phase transition corresponds to an ultraviolet fixed point of the continuum limit, then one would expect noncanonical finite size scaling since such a fixed point is widely expected to be non-Gaussian in nature.

The \emph{Ansatz} \eqref{naiveCL} dictates the following finite size scaling of discrete quantities at fixed spacetime $4$-volume $V_{4}$. A discrete quantity associated with the units $a^{p}$ scales towards the continuum limit as $N_{4}^{-p/4}$. In particular, the discrete time coordinate $\tau$ scales as
\begin{equation}
\tau\longrightarrow\tilde{\tau}=\frac{\tau}{N_{4}^{1/4}},
\end{equation}
and the discrete spatial $3$-volume $N_{3}^{\mathrm{SL}}(\tau)$ scales as
\begin{equation}
N_{3}^{\mathrm{SL}}(\tau)\longrightarrow\tilde{N}_{3}^{\mathrm{SL}}(\tilde{\tau})=\frac{N_{3}^{\mathrm{SL}}(\tau)}{N_{4}^{3/4}}.
\end{equation}
Since the scaled discrete time coordinate $\tilde{\tau}$ and the scaled discrete spatial $3$-volume $\tilde{N}_{3}^{\mathrm{SL}}(\tilde{\tau})$ are of course still dimensionless, these discrete quantities match onto the dimensionless time coordinate
\begin{equation}\label{dimensionlesstime4}
\tilde{t}=\frac{t}{V_{4}^{1/4}}
\end{equation}
and the dimensionless spatial $3$-volume 
\begin{equation}\label{dimensionlessvolume4}
\tilde{V}_{3}(\tilde{t})=\frac{V_{3}(t)}{V_{4}^{3/4}},
\end{equation} 
in the continuum limit. 

The finite size scaling \emph{Ansatz} \eqref{naiveCL} establishes the largest scale characterizing an ensemble of causal triangulations, namely $\ell_{V_{4}}=V_{4}^{1/4}$. Momentarily, I identify this scale with the de Sitter length $\ell_{\mathrm{dS}}$ of the ensemble average quantum geometry in phase C. Take note, however, that the scale $\ell_{V_{4}}$ is still defined in terms of the arbitrary lattice spacing $a$. I thus only know the ratio $\ell_{V_{4}}/a$ for any given ensemble of causal triangulations.

\subsubsection{Accounting for the discrete observables}\label{modelobservables}

Given the model defined by the action \eqref{MSM3action4} and the finite size scaling \emph{Ansatz} \eqref{naiveCL}, I now account for the phenomenology of the observables discussed in subsection \ref{observables}. 

\paragraph{$1$-point function of the discrete spatial volume}

The maximally symmetric extremum of the action \eqref{MSM3action4} with the constraint \eqref{MSM3action4c} is Euclidean de Sitter space for which
\begin{equation}\label{EdSvolprof}
V_{3}^{(\mathrm{EdS})}(t)=2\pi^{2}\ell_{\mathrm{dS}}^{3}\cos^{3}{\left(\frac{\sqrt{g_{tt}}t}{\ell_{\mathrm{dS}}}\right)}
\end{equation}
for the de Sitter length $\ell_{\mathrm{dS}}=\sqrt{3/\Lambda_{\mathrm{ren}}}$ (provided that $\Lambda_{\mathrm{ren}}>0$). One might reasonably expect that $V_{3}^{(\mathrm{EdS})}(t)$ closely models $\langle N_{3}^{\mathrm{SL}}(\tau)\rangle$ within the central accumulation: the ground state is typically the most symmetric state, and the path integral is typically dominated by the extremum of the classical action. 

\begin{figure}[!ht]
\centering
\includegraphics[scale=0.6]{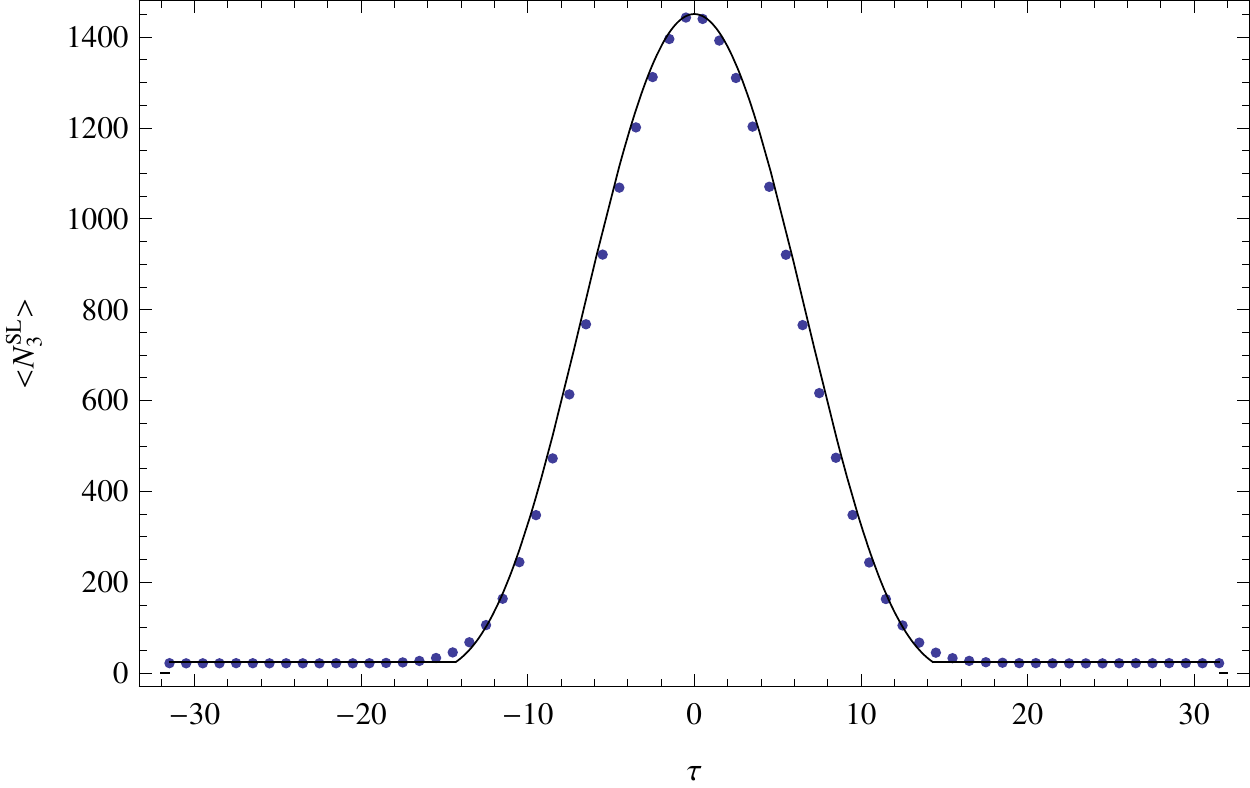}
\caption{The fit (in black) of the function \eqref{discreteEdSvolprof} to the coherent ensemble average number $\langle N_{3}^{\mathrm{SL}}\rangle$ of spacelike $3$-simplices as a function of the discrete time coordinate $\tau$ for an ensemble within phase C characterized by $\bar{T}=64$, $\bar{N}_{4}=81920$, $\kappa_{0}=2.0$, and $\Delta=0.4$.}
\label{fitvolprof3p1T64V80k}
\end{figure}

From the finite size scaling \emph{Ansatz} \eqref{naiveCL}, one can derive the discrete analogue $\mathcal{N}_{3}^{\mathrm{SL}}(\tau)$ of $V_{3}^{(\mathrm{EdS})}(t)$ following, for instance, \cite{JA&AG&JJ&RL2,JHC&JMM}. One finds that
\begin{equation}\label{discreteEdSvolprof}
\mathcal{N}_{3}^{\mathrm{SL}}(\tau)=\frac{3}{8}\frac{\langle N_{4}^{(1,4)}\rangle}{\bar{s}_{0}\langle N_{4}^{(1,4)}\rangle^{1/4}}\cos^{3}{\left(\frac{\tau}{\bar{s}_{0}\langle N_{4}^{(1,4)}\rangle^{1/4}}\right)}
\end{equation}
for the parameter
\begin{equation}
\bar{s}_{0}=\frac{2^{1/4}(1+\xi)^{1/4}\ell_{\mathrm{dS}}}{V_{4}^{1/4}\sqrt{g_{tt}}}
\end{equation}
with
\begin{equation}
\xi=\frac{\langle N_{4}^{(2,3)}\rangle+\langle N_{4}^{(3,2)}\rangle}{\langle N_{4}^{(1,4)}\rangle+\langle N_{4}^{(4,1)}\rangle}.
\end{equation}
With the spacetime $4$-volume $V_{4}$ explicitly identified as that of Euclidean de Sitter space, 
\begin{equation}\label{defs0}
\bar{s}_{0}=\frac{1}{\sqrt{g_{tt}}}\left[\frac{3(1+\xi)}{4\pi^{2}}\right]^{1/4}.
\end{equation}
The parameter $\bar{s}_{0}$ is proportional to (the inverse square root of) the real constant $g_{tt}$. The nature of $\bar{s}_{0}$ is thus laid bare: it sets the scale of the discrete time coordinate $\tau$ relative to the scale of the global time coordinate $t$. Just like $g_{tt}$, $\bar{s}_{0}$ is a gauge parameter; indeed, one could arrive at any value of $\bar{s}_{0}$ by choosing to label the time slices in a perverse manner. In figure \ref{fitvolprof3p1T64V80k} I display the fit of the function \eqref{discreteEdSvolprof} to $\langle N_{3}^{\mathrm{SL}}(\tau)\rangle$ within the central accumulation for a typical ensemble of causal triangulations within phase C. 
Clearly, the function \eqref{discreteEdSvolprof} provides a good fit of $\langle N_{3}^{\mathrm{SL}}(\tau)\rangle$. One may quantify the goodness of fit following, for instance, the method of \cite{JHC&JMM}. The fit yields the value of $\bar{s}_{0}$ not the value of $\ell_{\mathrm{dS}}$ (in units of $a$) because one measures $\tilde{t}$ in units of $\ell_{\mathrm{dS}}$. The fit thus only compares the shape of $V_{3}^{(\mathrm{EdS})}(t)$ to the shape of $\langle N_{3}^{\mathrm{SL}}(\tau)\rangle$.



With the ensemble average quantum geometry on large scales identified as that of Euclidean de Sitter space, I obtain the identification
\begin{equation}
V_{4}=\frac{8\pi^{2}\ell_{\mathrm{dS}}^{4}}{3}=C_{4}N_{4}a^{4}.
\end{equation}
The overall scale $\ell_{V_{4}}$ of the ensemble average quantum geometry---that of the central accumulation---is thus set by the de Sitter length $\ell_{\mathrm{dS}}$. Since Euclidean de Sitter space is a round sphere, its extent determines its duration. In particular, the de Sitter length in units of the lattice spacing is
\begin{equation}\label{desitterlengthlatticespacing}
\ell_{\mathrm{dS}}=\left[\frac{3C_{4}N_{4}}{8\pi^{2}}\right]^{1/4}a.
\end{equation}
The ratio $\ell_{\mathrm{dS}}/a$ depends not only on the number $N_{4}$, but also on the bare couplings $\kappa_{0}$ and $\Delta$ through the ratios $\langle N_{4}^{(p,q)}\rangle /N_{4}$ entering the expression \eqref{C4} for $C_{4}$. 

\paragraph{$2$-point function of fluctuations in the discrete spatial volume}\label{model2ptfunction}

Letting $V_{3}(t)=V_{3}^{(\mathrm{EdS})}(t)+v_{3}(t)$ for $V_{3}^{(\mathrm{EdS})}(t)\gg v_{3}(t)$, I expand the action \eqref{MSM3action4} to second order in $v_{3}(t)$, finding that
\begin{equation}\label{expandedaction}
S^{(\mathrm{E})}[v_{3}]=\frac{\pi \ell_{\mathrm{dS}}^{2}}{G_{\mathrm{ren}}}-\frac{1}{48\pi^{3}G_{\mathrm{ren}}\ell_{\mathrm{dS}}^{3}\sqrt{g_{tt}}}\int\mathrm{d}t\,v_{3}\left\{\frac{\mathrm{d}}{\mathrm{d}t}\left[\sec^{3}{\left(\frac{\sqrt{g_{tt}}t}{\ell_{\mathrm{dS}}}\right)}\frac{\mathrm{d}}{\mathrm{d}t}\right]+\frac{4g_{tt}}{\ell_{\mathrm{dS}}^{2}}\sec^{5}{\left(\frac{\sqrt{g_{tt}}t}{\ell_{\mathrm{dS}}}\right)}\right\}v_{3}+O(v_{3}^{3}).
\end{equation}
There are no terms of first order in $v_{3}(t)$ since I expand about an extremum of the action \eqref{MSM3action4}. Evidently, I am considering linearized perturbations $v_{3}(t)$ of the geometry of Euclidean de Sitter space. One may compute their connected $2$-point function $\langle v_{3}(t)v_{3}(t')\rangle$ by the path integration
\begin{equation}
\langle v_{3}(t)v_{3}(t')\rangle=\frac{\int\mathcal{D}v_{3}\,e^{-S^{(\mathrm{E})}[v_{3}]/\hbar}v_{3}(t)v_{3}(t')}{\int\mathcal{D}v_{3}\,e^{-S^{(\mathrm{E})}[v_{3}]/\hbar}}
\end{equation}
subject to the boundary conditions $v_{3}\left(-\pi \ell_{\mathrm{dS}}/2\sqrt{g_{tt}}\right)=0$ and $v_{3}\left(+\pi \ell_{\mathrm{dS}}/2\sqrt{g_{tt}}\right)=0$ and the constraint $\int\mathrm{d}t\sqrt{g_{tt}}v_{3}=0$. A standard calculation demonstrates that $\langle v_{3}(t)v_{3}(t')\rangle$ is given by the inverse of the differential operator
\begin{equation}\label{differentialoperator}
-\frac{1}{48\pi^{3}\hbar G_{\mathrm{ren}}\ell_{\mathrm{dS}}^{3}\sqrt{g_{tt}}}\left\{\frac{\mathrm{d}}{\mathrm{d}t}\left[\sec^{3}{\left(\frac{\sqrt{g_{tt}}t}{\ell_{\mathrm{dS}}}\right)}\frac{\mathrm{d}}{\mathrm{d}t}\right]+\frac{4g_{tt}}{\ell_{\mathrm{dS}}^{2}}\sec^{5}{\left(\frac{\sqrt{g_{tt}}t}{\ell_{\mathrm{dS}}}\right)}\right\}
\end{equation}
appearing in the term of second order in $v_{3}$ in equation \eqref{expandedaction}. One can decompose $\langle v_{3}(t)v_{3}(t')\rangle$ as
\begin{equation}
\langle v_{3}(t)v_{3}(t')\rangle=\sum_{j}\frac{1}{\upsilon_{j}}\left[\nu_{3}(t)\right]_{j}\left[\nu_{3}(t')\right]_{j}
\end{equation}
for the eigenfunctions $\left[\nu_{3}(t)\right]_{j}$ and associated eigenvalues $\upsilon_{j}$ solving the characteristic equation
\begin{equation}\label{characteristicequation}
-\frac{1}{48\pi^{3}\hbar G_{\mathrm{ren}}\ell_{\mathrm{dS}}^{3}\sqrt{g_{tt}}}\left\{\frac{\mathrm{d}}{\mathrm{d}t}\left[\sec^{3}{\left(\frac{\sqrt{g_{tt}}t}{\ell_{\mathrm{dS}}}\right)}\frac{\mathrm{d}}{\mathrm{d}t}\right]+\frac{4g_{tt}}{\ell_{\mathrm{dS}}^{2}}\sec^{5}{\left(\frac{\sqrt{g_{tt}}t}{\ell_{\mathrm{dS}}}\right)}\right\}\left[\nu_{3}(t)\right]_{j}=\upsilon_{j}\left[\nu_{3}(t)\right]_{j}.
\end{equation}
The eigenvalues $\upsilon_{j}$ have the dimensions of an effective spacetime energy-momentum density. 

\begin{figure}[!ht]
\centering
\includegraphics[width=\linewidth]{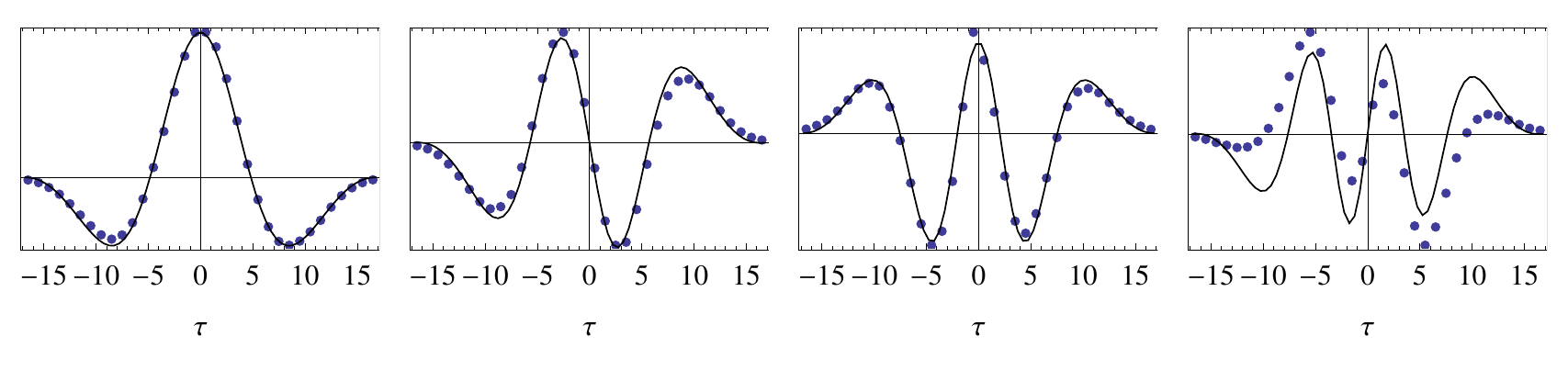}
\caption{The fit (in black) of the first four eigenfunctions $\left[\nu_{3}(t)\right]_{j}$ to the first four eigenvectors $[\eta_{3}^{\mathrm{SL}}(\tau)]_{j}$ of the connected $2$-point function $\langle n_{3}^{\mathrm{SL}}(\tau)n_{3}^{\mathrm{SL}}(\tau')\rangle$ within the central accumulation for an ensemble within phase C characterized by $\bar{T}=64$, $\bar{N}_{4}=81920$, $\kappa_{0}=2.0$, and $\Delta=0.4$.}
\label{eigenvectorfits}
\end{figure}

\begin{figure}[!ht]
\centering
\includegraphics[scale=0.6]{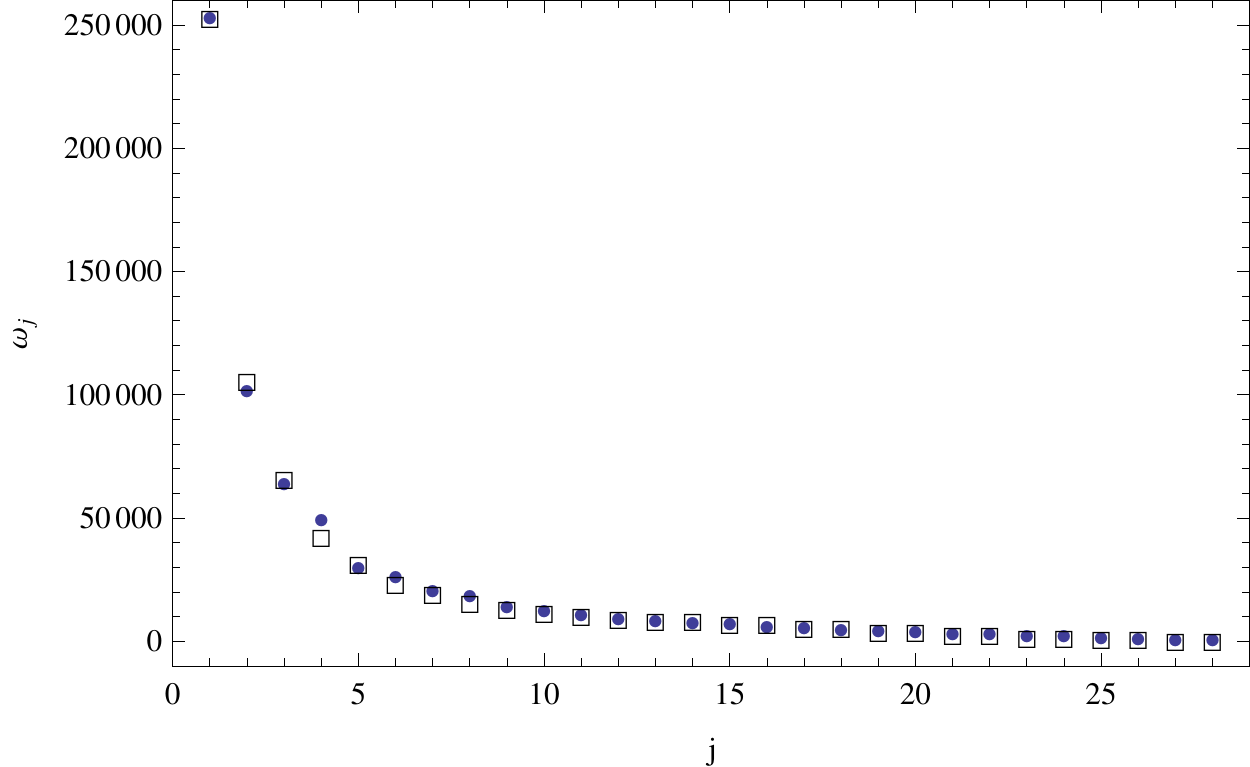}
\caption{The fit (in black boxes) of the inverse of the eigenvalues $\upsilon_{j}$ to the eigenvalues $\omega_{j}$ of the connected $2$-point function $\langle n_{3}^{\mathrm{SL}}(\tau)n_{3}^{\mathrm{SL}}(\tau')\rangle$ within the central accumulation for an ensemble within phase C characterized by $\bar{T}=64$, $\bar{N}_{4}=81920$, $\kappa_{0}=2.0$, and $\Delta=0.4$.}
\label{eigenvaluefit}
\end{figure}

The $2$-point function $\langle v_{3}(t)v_{3}(t')\rangle$ thus defines a ladder of scales corresponding to its eigenvalue spectrum. I now connect the eigenvalue spectrum $\upsilon_{j}$ to the numerically measured eigenvalue spectrum $\omega_{j}$ again using the finite size scaling \emph{Ansatz} \eqref{naiveCL}. The discrete analogue of equation \eqref{characteristicequation} is
\begin{eqnarray}\label{discretecharacteristicequation}
\lefteqn{-\frac{1}{48\pi^{3}\hbar G_{\mathrm{ren}}\ell_{\mathrm{dS}}^{3}\sqrt{g_{tt}}}\left(\frac{V_{4}}{N_{4}}\right)^{5/4}\Bigg\{\frac{\mathrm{d}}{\mathrm{d}\tau}\left[\sec^{3}{\left(\frac{\tau}{\bar{s}_{0}\langle N_{4}^{(1,4)}\rangle^{1/4}}\right)}\frac{\mathrm{d}}{\mathrm{d}\tau}\right]}\nonumber\\ && \qquad\qquad\qquad\qquad+\left(\frac{2}{\bar{s}_{0}\langle N_{4}^{(1,4)}\rangle^{1/4}}\right)^{2}\sec^{5}{\left(\frac{\tau}{\bar{s}_{0}\langle N_{4}^{(1,4)}\rangle^{1/4}}\right)}\Bigg\}\left[\eta_{3}^{\mathrm{SL}}(\tau)\right]_{j}=\frac{1}{\omega}_{j}\left[\eta_{3}^{\mathrm{SL}}(\tau)\right]_{j},
\end{eqnarray}
for a discretization of the derivative with respect to the discrete time coordinate $\tau$. The inverse of the eigenvalue $\omega_{j}$ appears in equation \eqref{discretecharacteristicequation} because the $2$-point function is the inverse of the differential operator \eqref{differentialoperator}. The predicted relation between the eigenvalues $\upsilon_{j}$ and the eigenvalues $\omega_{j}$ is thus
\begin{equation}\label{eigenvaluerelation}
\upsilon_{j}=\frac{1}{48\pi^{3}\hbar G_{\mathrm{ren}}\ell_{\mathrm{dS}}^{3}\sqrt{g_{tt}}}\left(\frac{V_{4}}{N_{4}}\right)^{5/4}\frac{1}{\omega}_{j}
\end{equation}
for a discretization of the differential operator \eqref{differentialoperator} such that there are $T_{\mathrm{ph}}$ eigenvalues $\upsilon_{j}$. In figure \ref{eigenvectorfits} I display the fit of the first four eigenfunctions $\left[\nu_{3}(t)\right]_{j}$ to the first four eigenvectors $[\eta_{3}^{\mathrm{SL}}(\tau)]_{j}$ for a typical ensemble of causal triangulations within phase C.
The fits in figure \ref{eigenvectorfits} do not involve any adjustable parameters: one simply computes the eigenfunctions $\left[\nu_{3}(t)\right]_{j}$ for the background value of $\bar{s}_{0}$ determined by the fit of figure \ref{fitvolprof3p1T64V80k}. Clearly, the model provides a better description of the eigenvectors $\left[\eta_{3}^{\mathrm{SL}}(\tau)\right]_{j}$ for low $j$. This concurs with the fact that the eigenvectors $\left[\eta_{3}^{\mathrm{SL}}(\tau)\right]_{j}$ for low $j$ represent larger scale fluctuations of the ensemble average quantum geometry. In figure \ref{eigenvaluefit} I display the fit of (the inverse of) the eigenvalues $\upsilon_{j}$ to the eigenvalues $\omega_{j}$ for the same ensemble.
The fit in figure \ref{eigenvaluefit} also does not involve any adjustable parameters: one simply computes the eigenvalues $\upsilon_{j}$ of the eigenfunctions $\left[\nu_{3}(t)\right]_{j}$. 

From the relation \eqref{eigenvaluerelation}, the definition \eqref{defs0} of $\bar{s}_{0}$, and the expression  \eqref{desitterlengthlatticespacing} for the de Sitter length in units of the lattice spacing, one can determine $\hbar G_{\mathrm{ren}}$ in units of the lattice spacing $a$ as
\begin{equation}\label{renG}
\hbar G_{\mathrm{ren}}=\frac{2^{7/4}\pi\bar{s}_{0}}{3^{3}C_{4}^{1/2}N_{4}^{7/4}(1+\xi)^{1/4}}\frac{1}{\upsilon_{j}\omega_{j}}a^{2}
\end{equation}
or in units of the de Sitter length $\ell_{\mathrm{dS}}$ as
\begin{equation}
\hbar G_{\mathrm{ren}}=\frac{2^{1/4}\bar{s}_{0}}{3^{5/2}N_{4}^{5/4}(1+\xi)^{1/4}}\frac{1}{\upsilon_{j}\omega_{j}}\ell_{\mathrm{dS}}^{2}.
\end{equation}
$\hbar G_{\mathrm{ren}}$ is of course the square of the Planck length $\ell_{\mathrm{P}}$, and the ratio $\hbar G_{\mathrm{ren}}/\ell_{\mathrm{dS}}^{2}$ is proportional to $G_{\mathrm{ren}}\Lambda_{\mathrm{ren}}$. 
With $j\in\{1,\ldots,T_{\mathrm{ph}}\}$ one obtains $T_{\mathrm{ph}}$ determinations of $\hbar G_{\mathrm{ren}}$ in either units. The model of $\langle n_{3}^{\mathrm{SL}}(\tau)n_{3}^{\mathrm{SL}}(\tau')\rangle$ assumes that $G_{\mathrm{ren}}$ has a definite value or, equivalently, that $G_{\mathrm{ren}}$ does not have a significant renormalization group flow on the applicable scales. Accordingly, I average the $T_{\mathrm{ph}}$ determinations of $\hbar G_{\mathrm{ren}}$. The variance in the $T_{\mathrm{ph}}$ determinations of $\hbar G_{\mathrm{ren}}/a^{2}$ quantifies to some extent the validity of this assumption. Since the eigenvectors $\left[\eta_{3}^{\mathrm{SL}}(\tau)\right]_{j}$ probe the quantum geometry on successively smaller scales, one could in principle employ a refined model in which the eigenvalues $\omega_{j}$ yield the renormalization group flow of $G_{\mathrm{ren}}$. This analysis closely follows that of \cite{JA&AG&JJ&RL2}.



\paragraph{Spectral dimension}

The model defined by the action \eqref{MSM3action4} does not obviously apply to the numerical measurements of the spectral dimension discussed in subsubsection \ref{SpectralDimension}. Recall, however, two facts: the spectral dimension provides a scale-dependent measure of the effective dimensionality, and this model pertains to the ensemble average quantum geometry on large scales. One might thus expect this model to apply to the spectral dimension for sufficiently long diffusion times. Benedetti and Henson have tested this expectation in the case of $2+1$ dimensions as I now explain \cite{DB&JH}.

On a $(d+1)$-dimensional Riemannian manifold $\mathscr{M}$ of metric tensor $\mathbf{g}$ for coordinates $x$, the heat equation takes the form
\begin{equation}\label{continuousheatequation}
\frac{\partial}{\partial u}K(x,x',u)+\nabla^{2}K(x,x',u)=0
\end{equation}
for the diffusion time $u$ and the Laplacian operator $\nabla^{2}$ of the metric tensor $\mathbf{g}$. Subject to the initial condition
\begin{equation}
K(x,x',0)=\frac{1}{\sqrt{g(x')}}\delta^{(d+1)}(x'-x),
\end{equation}
one may express the heat kernel as
\begin{equation}\label{heatkernel}
K(x,x',u)=\sum_{j}e^{-\varsigma_{j}u}\zeta_{j}(x)\zeta_{j}(x')
\end{equation}
for the eigenfunctions $\zeta_{j}(x)$ and associated eigenvalues $\varsigma_{j}$ of the Laplacian operator. The return probability is computed as
\begin{equation}\label{defreturnprobability}
P(u)=\frac{1}{V_{d+1}}\int_{\mathscr{M}}\mathrm{d}^{d+1}x\sqrt{g(x)}K(x,x,u)
\end{equation}
for the $(d+1)$-volume
\begin{equation}
V_{d+1}=\int_{\mathscr{M}}\mathrm{d}^{d+1}x\sqrt{g(x)}
\end{equation}
of the manifold $\mathscr{M}$. 
Substituting the expression \eqref{heatkernel} into the definition \eqref{defreturnprobability} of the return probability, one finds that
\begin{equation}\label{heattrace}
P(u)=\frac{1}{V_{d+1}}\sum_{j}e^{-\varsigma_{j}u}.
\end{equation}
The spectral dimension is then
\begin{equation}\label{specdimeigenvalues}
d_{\mathfrak{s}}(u)=2u\frac{\sum_{j}\varsigma_{j}e^{-\varsigma_{j}u}}{\sum_{j}e^{-\varsigma_{j}u}},
\end{equation}
determined completely by the eigenvalues $\varsigma_{j}$. 

\begin{figure}[ht!]
\centering
\includegraphics[scale=0.65]{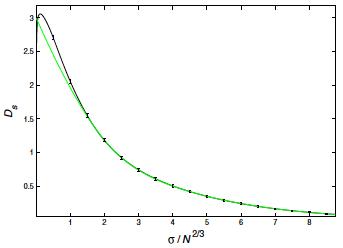}
\caption{The fit (in green) of the ensemble average spectral dimension $\langle d_{s}\rangle$ as a function of finite size scale diffusion time $\sigma/N_{3}^{2/3}$ for an ensemble of $(2+1)$-dimensional causal triangulations within phase C.}
\label{specdimfit}
\end{figure}

Benedetti and Henson computed the eigenvalues of the Laplacian operator for the metric tensor $\mathbf{g}$ specified by the line element
\begin{equation}\label{stretchedspheremetric}
\mathrm{ds}^{2}=g_{tt}\mathrm{d}t^{2}+\ell_{\mathrm{dS}}^{2}\cos^{2}{\left(\frac{t}{\ell_{\mathrm{dS}}}\right)}\left(\mathrm{d}\theta^{2}+\sin^{2}{\theta}\,\mathrm{d}\varphi^{2}\right).
\end{equation}
The line element \eqref{stretchedspheremetric} describes a $(2+1)$-dimensional stretched sphere, which coincides with Euclidean de Sitter space for $g_{tt}=1$. From these eigenvalues they computed the spectral dimension according to equation \eqref{specdimeigenvalues}. They then performed a fit of this spectral dimension to the ensemble average spectral dimension $\langle d_{\mathfrak{s}}(\sigma)\rangle$ for sufficiently large diffusion times $\sigma$ as depicted in figure \ref{specdimfit}. 
This fit involves identifying the dimensionless continuous diffusion time $\tilde{u}=u/V_{3}^{2/3}$ with the finite size scaled discrete diffusion time $\tilde{\sigma}=\sigma/N_{3}^{2/3}$. 
This fit also involves determining the diffusion time $\sigma_{\mathrm{cos}}$, in excess of $\sigma_{\mathrm{cl}}$, at which $\langle d_{\mathfrak{s}}(\sigma)\rangle$ begins to exhibit stretched sphere behavior. They found the spectral dimension of the stretched sphere to provide a good fit of $\langle d_{\mathfrak{s}}(\sigma)\rangle$ on these scales. The fit does not yield $\ell_{\mathrm{dS}}$ (in units of $a$) because one measures $u$ in units of $\ell_{\mathrm{dS}}$. The fit thus only compares the shape of the stretched sphere's spectral dimension to the shape of $\langle d_{\mathfrak{s}}(\sigma)\rangle$ on sufficiently large scales. To obtain a measurement of $\ell_{\mathrm{dS}}$ in units of the lattice spacing $a$ from $\langle d_{\mathfrak{s}}(\sigma)\rangle$, one can employ the method reported in \cite{JHC2} for putting the diffusion time $\sigma$ in contact with the lattice spacing $a$.\footnote{One might wonder why Benedetti and Henson chose to fit the spectral dimension of the stretched sphere, not that of Euclidean de Sitter space, to $\langle d_{\mathfrak{s}}(\sigma)\rangle$. Their motivation was twofold: first, they pointed out that the random walker probes the quantum geometry more generally than does the observable $N_{2}^{\mathrm{SL}}(\tau)$, and, second, they were interested in possibly finding evidence for a Ho\v{r}ava-Lifshitz-like continuum limit as suggested in \cite{CDTandHL}. The stretched sphere is not a solution of the Einstein equations except for $g_{tt}=1$ but is a solution of the Ho\v{r}ava-Lifshitz equations. This second point suggests that their use of the finite size scaling \emph{Ansatz} \eqref{naiveCL} is not necessarily compatible with their choice of the line element \eqref{stretchedspheremetric} since the finite size scaling \emph{Ansatz} assumes isotropic scaling.}

For the line element \eqref{lineelement} with the scale factor of Euclidean de Sitter space, the eigenvalues of the Laplacian are
\begin{equation}
\varsigma_{jkl}=[\beta(j,l)+k-(j-l)][\beta(j,l)+k-(j-l)+1]-2
\end{equation}
with
\begin{equation}
\beta^{2}(j,l)=\left[\sqrt{l(l+1)+\frac{1}{4}}+j-l\right]\left[\sqrt{l(l+1)+\frac{1}{4}}+j-l+1\right]+\frac{1}{4}
\end{equation}
for $j\geq1$, $0\leq l\leq j$, and $0\leq k\leq j-l$. Each eigenvalue $\varsigma_{jkl}$ has multiplicity $2l+1$. Given numerical measurements of the spectral dimension for sufficiently large diffusion times, one could thus repeat the analysis of Benedetti and Henson for the case of $3+1$ dimensions. In combination with the method presented in \cite{JHC2}, one could independently determine $\ell_{\mathrm{dS}}$ in units of $a$.

\section{A renormalization group scheme}\label{renormalization}

I now propose a renormalization group scheme for causal dynamical triangulations. Its aim is the extraction of the renormalization group flows of the continuum limit's couplings. This scheme applies to the portion of phase C for which the model of subsection \ref{model} provides a good description of the observables of subsection \ref{observables}. Once one has further studied the remaining portions of phase C, one can extend the renormalization group scheme following the general method of paper I. 
Hopefully, the scheme proposed below serves to guide such extensions.

\subsection{Expectations}\label{guidelines}


Before presenting the renormalization group scheme in detail, I wish to formulate some general expectations for the effects of renormalization on the observables of subsection \ref{observables}. I  base these expectations on general considerations of the scale-dependent dynamics present within phase C. I intend these expectations to serve as guideposts along the path to the complete renormalization group scheme. 

Consider the typical case of a renormalization group transformation that incrementally increases the ultraviolet scale $\ell_{\mathrm{UV}}$ and leaves fixed the infrared scale $\ell_{\mathrm{IR}}$. What again is the physical import of such a renormalization group transformation? This transformation integrates out degrees of freedom on the interval of scales $(\ell_{\mathrm{UV}},\ell_{\mathrm{UV}}+\delta\ell)$, imprinting their dynamics on the resulting renormalization of the couplings. Once this transformation has acted, the degrees of freedom on the interval of scales $(\ell_{\mathrm{UV}},\ell_{\mathrm{UV}}+\delta\ell)$ are of course no longer accessible. 
How do I expect this transformation to affect the three observables discussed in section \ref{phenomenology}? 

The discrete spatial $3$-volume $N_{3}^{\mathrm{SL}}(\tau)$ as a function of the discrete time coordinate only probes the quantum geometry on its largest scales. As the renormalization group transformation only affects the quantum geometry on its smallest scales, I do not expect any effect on $N_{3}^{\mathrm{SL}}(\tau)$. More precisely, I expect the shape of the ensemble average $\langle N_{3}^{\mathrm{SL}}(\tau)\rangle$ to remain equally well-described as the shape of $V_{3}^{(\mathrm{EdS})}(t)$, the spatial $3$-volume of Euclidean de Sitter space. 
What does this expectation imply for the parameter $\bar{s}_{0}$ characterizing the function \eqref{discreteEdSvolprof}, the discrete analogue of $V_{3}^{\mathrm{(EdS)}}(t)$ fit to $\langle N_{3}^{\mathrm{SL}}(\tau)\rangle$? As I explained previously, $\bar{s}_{0}$ is a gauge parameter relating the scale of the discrete time coordinate $\tau$ to the scale of the continuous time coordinate $t$. As such the value of $\bar{s}_{0}$ is not physically relevant, so this expectation allows for $\bar{s}_{0}$ to change under a renormalization group transformation. 

The connected $2$-point function $n_{3}^{\mathrm{SL}}(\tau)n_{3}^{\mathrm{SL}}(\tau')$ of fluctuations in the discrete spatial $3$-volume 
probes the quantum geometry on multiple scales. The eigenvectors $\left[\eta_{3}^{\mathrm{SL}}(\tau)\right]_{j}$, into which I previously decomposed the ensemble average $\langle n_{3}^{\mathrm{SL}}(\tau)n_{3}^{\mathrm{SL}}(\tau')\rangle$, represent excitations of the quantum geometry on successively smaller scales. As the renormalization group transformation incrementally integrates out degrees of freedom, I expect $\langle n_{3}^{\mathrm{SL}}(\tau)n_{3}^{\mathrm{SL}}(\tau')\rangle$ to decompose into incrementally fewer eigenvectors $\left[\eta_{3}^{\mathrm{SL}}(\tau)\right]_{j}$.
Since the number $T_{\mathrm{ph}}$ of time slices within the central accumulation determines the number of nontrivial eigenvectors $\left[\eta_{3}^{\mathrm{SL}}(\tau)\right]_{j}$, I expect $T_{\mathrm{ph}}$ to decrease incrementally. What does this expectation imply for the eigenvalues $\omega_{j}$ of the eigenvectors $\left[\eta_{3}^{\mathrm{SL}}(\tau)\right]_{j}$? The eigenvalues $\omega_{j}$ enter into the determination of the renormalized Newton constant $G_{\mathrm{ren}}$ in units of the lattice spacing $a$. Without making any \emph{a priori} assumptions regarding the renormalization group flow of $G_{\mathrm{ren}}$, I cannot determine how the eigenvalues $\omega_{j}$ might change. 

The spectral dimension $d_{\mathfrak{s}}(\sigma)$ as a function of the diffusion time also probes the quantum geometry on multiple scales. 
On sufficiently small scales the ensemble average $\langle d_{\mathfrak{s}}(\sigma)\rangle$ exhibits dynamical dimensional reduction. This phenomenon manifestly results from the dynamics of degrees of freedom on these scales. As the renormalization group transformation incrementally integrates out degrees of freedom, I expect the dimensional reduction to become incrementally less evident. What does this expectation imply for the scales $\ell_{\mathrm{qm}}$ and $\ell_{\mathrm{cl}}$ characterizing the behavior of $\langle d_{\mathfrak{s}}(\sigma)\rangle$? I expect the scale $\ell_{\mathrm{qm}}$ to become incrementally more difficult to extract as it 
eventually falls outside of the interval of scales $(\ell_{\mathrm{UV}},\ell_{\mathrm{IR}})$. I expect the scale $\ell_{\mathrm{cl}}$ to correspond to incrementally smaller values of the diffusion time $\sigma$ as each step in the diffusion process corresponds to an incrementally larger scale. 
On sufficiently large scales the ensemble average $\langle d_{\mathfrak{s}}(\sigma)\rangle$ exhibits an exponential decay set by the de Sitter length $\ell_{\mathrm{dS}}$. 
Since a renormalization group transformation only affects small scales, I do not expect any effect on the decay of $\langle d_{\mathfrak{s}}(\sigma)\rangle$. More specifically, I expect the shape of $\langle d_{\mathfrak{s}}(\sigma)\rangle$ to remain equally well-described as the shape of $d_{\mathfrak{s}}(u)$ for Euclidean de Sitter space on these scales. 

Any renormalization group scheme for phase C based on a transformation that incrementally increases $\ell_{\mathrm{UV}}$ and leaves fixed $\ell_{\mathrm{IR}}$ should meet these expectations. These expectations furnish criteria by which to judge a proposed renormalization group scheme. Alternatively, these expectations serve to supplement a renormalization group scheme that does not otherwise completely delineate its own renormalization group trajectories. 
Given the relative paucity of physical observables that one currently knows how to extract by numerical measurement from phase C, these expectations likely assume the latter role in the renormalization group scheme that I now propose. 


\subsection{Formulation}\label{formulation}

I reviewed in subsection \ref{recap} the key steps in the formulation and implementation of a renormalization group scheme for a lattice-regularized quantum theory of gravity as developed in paper I. 
I now explain how each of these steps is realized in my scheme. 

\subsubsection{Selection of a model for the continuum limit}


I choose as a model that defined by the effective action \eqref{MSM3action4} subject to the constraint \eqref{MSM3action4c}, a minisuperspace approximation of the Einstein-Hilbert truncation of the gravitational effective action, 
Wick rotated to the Euclidean sector.
Accordingly, I am interested in the renormalization group flows of the renormalized Newton constant $G_{\mathrm{ren}}$ and the renormalized cosmological constant $\Lambda_{\mathrm{ren}}$ (assuming \emph{a priori} that both of these couplings flow). In light of the discussion of subsection \ref{model}, the action \eqref{MSM3action4} evidently provides a good description of the observables that I intend to measure on the length scales to which these measurements have access. 


\subsubsection{Selection of a finite size scaling \emph{Ansatz}}

I choose to employ the canonical finite size scaling \emph{Ansatz} \eqref{naiveCL}. The \emph{Ansatz} \eqref{naiveCL} is consistent with my choice of model: the quantity on which the \emph{Ansatz} \eqref{naiveCL} is based---the spacetime $4$-volume $V_{4}$---is a physical observable of the model. In light of the discussion of subsection \ref{model}, the \emph{Ansatz} \eqref{naiveCL} evidently provides a good description of the finite size scaling towards the continuum limit of the observables that I intend to measure on the length scales to which these measurements have access.The \emph{Ansatz} \eqref{naiveCL} might not apply to $\langle d_{\mathfrak{s}}(\sigma)\rangle$ for $\sigma<\sigma_{\mathrm{cl}}$, but I do not require its applicability for my renormalization group scheme.

\subsubsection{Delineation of a renormalization group trajectory}


The delineation of a renormalization group trajectory relies firstly on the identification of physical observables of the chosen model, specifically physical observables that one can infer \emph{via} the finite size scaling \emph{Ansatz} from numerical measurements on an ensemble of causal triangulations. 
With respect to the chosen model, all physical observables are geometric invariants constructed from integrals involving the metric tensor and its derivatives over a spacetime region. Of the several quantities that I discussed in subsection \ref{model}, which constitute physical observables of this model? 

The spacetime $4$-volume $V_{4}$ is a physical observable. $V_{4}$ is moreover scale-independent and dimensionful. Measured with respect to a physical standard unit of length $\ell_{\mathrm{unit}}$, $V_{4}$ is therefore just a positive real number whose value should not change under a renormalization group transformation. The chosen finite size scaling \emph{Ansatz} dictates $V_{4}$ in units of the lattice spacing $a$. Since the lattice spacing is arbitrary, one must determine a physical standard unit of length in terms of which one measures $V_{4}$. I discuss the choice of a standard unit of length below. 

The spatial $3$-volume $V_{3}(t)$ as a function of the global time coordinate---the continuum limit of $\langle N_{3}^{\mathrm{SL}}(\tau)\rangle$---is not itself a physical observable since $V_{3}(t)$ explicitly depends on a choice of time coordinate. $V_{3}(t)$ does nevertheless contain some physical information: one infers the Euclidean de Sitter form of the large scale ensemble average quantum geometry by matching the shape of  $V_{3}^{(\mathrm{EdS})}(t)$ to the shape of $\langle N_{3}^{\mathrm{SL}}(\tau)\rangle$. The connected $2$-point function $\langle v_{3}(t)v_{3}(t')\rangle$ of fluctuations in the spatial $3$-volume---the continuum limit of $\langle n_{3}^{\mathrm{SL}}(\tau)n_{3}^{\mathrm{SL}}(\tau')\rangle$---is also not itself a physical observable for the same reason. More specifically, the eigenvalues $\upsilon_{j}$ of the eigenfunctions $\left[\nu_{3}(t)\right]_{j}$ of $\langle v_{3}(t)v_{3}(t')\rangle$---the continuum limits of the eigenvalues $\omega_{j}$ of the eigenvectors $\left[\eta_{3}^{\mathrm{SL}}(\tau)\right]_{j}$ of $\langle n_{3}^{\mathrm{SL}}(\tau)n_{3}^{\mathrm{SL}}(\tau')\rangle$---are not physical observables either since $\upsilon_{j}$ explicitly depends on the gauge parameter $g_{tt}$. $\langle v_{3}(t)v_{3}(t')\rangle$ does nevertheless contain some physical information: one infers the Euclidean de Sitter form of fluctuations in the large scale ensemble average quantum geometry by matching the shape of $\langle v_{3}(t)v_{3}(t')\rangle$ to the shape of $\langle n_{3}^{\mathrm{SL}}(\tau)n_{3}^{\mathrm{SL}}(\tau')\rangle$. One only gathers this information by first hypothesizing that the discrete time coordinate $\tau$ has as its continuum limit the global time coordinate $t$. 

The spectral dimension $d_{\mathfrak{s}}(u)$ as a function of the diffusion time---the continuum limit of $\langle d_{\mathfrak{s}}(\sigma)\rangle$---is a physical observable. Even though the chosen model does not describe the dynamical dimensional reduction of $\langle d_{\mathfrak{s}}(\sigma)\rangle$ on sufficiently small scales, the spectral dimension on these and all other scales is constructed in the manner of physical observables of the chosen model. Clearly, $d_{\mathfrak{s}}(u)$ is scale-dependent and dimensionless. 

Of the two physical observables that I just identified---the spacetime $4$-volume $V_{4}$ and the spectral dimension $d_{\mathfrak{s}}(u)$---which are viable for the delineation of a renormalization group trajectory? As a scale-independent and dimensionful physical observable, $V_{4}$ is viable given a choice of standard unit of length $\ell_{\mathrm{unit}}$. The value of $V_{4}$ in units of $\ell_{\mathrm{unit}}$ should then be constant along a renormalization group trajectory. As a scale-dependent and dimensionless physical observable, which one can moreover measure over essentially the whole interval of scales $(\ell_{\mathrm{UV}},\ell_{\mathrm{IR}})$, $d_{\mathfrak{s}}(u)$ is in principle supremely viable. The value of $d_{\mathfrak{s}}(u)$ at a particular diffusion time, which corresponds to some particular scale, should be constant along a renormalization group trajectory (until that scale is no longer accessible). Currently, one only understands the connection between $\langle d_{\mathfrak{s}}(\sigma)\rangle$ and $d_{\mathfrak{s}}(u)$ for $\sigma\in(\sigma_{\mathrm{cos}},\infty)$. For these sufficiently large discrete diffusion times, the finite size scaling \emph{Ansatz} \eqref{naiveCL} obtains, mapping $\sigma$ into $u$ \emph{via} the identification of $\tilde{\sigma}=\sigma/N_{4}^{1/2}$ with $\tilde{u}=u/V_{4}^{1/2}$. Accordingly, the value of $\langle d_{\mathfrak{s}}(\sigma(u))\rangle$ at any particular value of $u\in(u_{\mathrm{cos}},\infty)$ should be constant along a renormalization group trajectory. 

To complete the prescription for delineating renormalization group trajectories, I now consider the candidates for a standard unit of length $\ell_{\mathrm{unit}}$ in terms of which all other length scales are measured. I explore the consequences for my renormalization group scheme of each such choice. I find that some choices are not viable while other choices are viable. My renormalization group scheme thus contains a residual ambiguity: the choice of a standard unit of length. 
I hope that an application of my scheme to numerically simulated ensembles of causal triangulations within phase C resolves this ambiguity. 
More than one choice may prove physically consistent, a possibility potentially indicative of multiple different choices for a standard unit of length being mutually consistent. 

\paragraph{Lattice spacing $a$ as the standard unit of length $\ell_{\mathrm{unit}}$}

Consider using the lattice spacing $a$ as the standard unit of length $\ell_{\mathrm{unit}}$. Although I have continually advised against making such a choice, I wish to consider its implications if only to reinforce my point. 
Constancy of the spacetime $4$-volume, measured in units of the lattice spacing $a$, dictates  \emph{via} equation \eqref{naiveCL} constancy of the product $C_{4}N_{4}$. 
Accordingly, one must compensate for any change in the number $N_{4}$ of $4$-simplices by a commensurate change in the discrete $4$-volume $C_{4}$ of an effective $4$-simplex. Since $C_{4}$ depends on the bare couplings $\kappa_{0}$ and $\Delta$ but varies only weakly with $N_{4}$, one cannot change $N_{4}$ while holding $\kappa_{0}$ and $\Delta$ fixed. This is the only definite condition on the renormalization group trajectories that follows from this choice for $\ell_{\mathrm{unit}}$. Whether or not one can maintain this condition is a test of this choice. 

Although I can compute $\ell_{\mathrm{P}}/a$ and in principle $\ell_{\mathrm{cl}}/a$ and $\ell_{\mathrm{qm}}/a$, I do not know how these quantities should change along a renormalization group trajectory beyond the restrictions imposed by the general expectations of subsection \ref{guidelines}. One could impose these expectations as further conditions on the renormalization group trajectories. Note that the interval of scales $(\tilde{\ell}_{\mathrm{UV}},\tilde{\ell}_{\mathrm{IR}})$ is invariant along a renormalization group trajectory for this choice of standard unit of length. $\tilde{\ell}_{\mathrm{UV}}$ is unity by definition, and $\tilde{\ell}_{\mathrm{IR}}$ is proportional to the fourth root of $C_{4}N_{4}$. There is thus no clear parameter to parametrize the renormalization group flow. 


\paragraph{de Sitter length $\ell_{\mathrm{dS}}$ as the standard unit of length $\ell_{\mathrm{unit}}$}

Consider using the de Sitter length $\ell_{\mathrm{dS}}$ as the standard unit of length $\ell_{\mathrm{unit}}$. Constancy of the spacetime $4$-volume $V_{4}$, measured in units of the de Sitter length $\ell_{\mathrm{dS}}$, now holds by definition upon identification of $V_{4}$ as that of Euclidean de Sitter space. 
This choice for the standard unit of length thus trivializes the condition of constant $V_{4}$ along a renormalization group trajectory. 
Without any further conditions determining the renormalization group trajectories, there is thus no way to define them. Within the renormalization group scheme that I propose, this choice of standard unit of length simply does not work in its desired capacity.

\paragraph{Planck length $\ell_{\mathrm{P}}$ as the standard unit of length $\ell_{\mathrm{unit}}$}

Consider using the Planck length $\ell_{\mathrm{P}}$ as the standard unit of length $\ell_{\mathrm{unit}}$. Constancy of the spacetime $4$-volume $V_{4}$, measured in units of the Planck length $\ell_{\mathrm{P}}$, dictates constancy of the ratio $\ell_{\mathrm{dS}}/\ell_{\mathrm{P}}$ upon identification of $V_{4}$ as that of Euclidean de Sitter space. 
Equivalently, the dimensionless product $G_{\mathrm{ren}}\Lambda_{\mathrm{ren}}$ remains constant along a renormalization group trajectory. Since $G_{\mathrm{ren}}$ and $\Lambda_{\mathrm{ren}}$ are the only couplings of the chosen model, there are no renormalization group flows for this choice of standard unit of length. The ratio $\ell_{\mathrm{dS}}/\ell_{\mathrm{P}}$ determines the magnitude of the fluctuations $v_{3}(t)$ in the spatial $3$-volume $V_{3}(t)$ relative to $V_{3}(t)$ itself. Accordingly, the magnitude of the fluctuations $n_{3}^{\mathrm{SL}}(\tau)$ in the number $N_{3}^{\mathrm{SL}}(\tau)$ of spacelike $3$-simplices relative to $N_{3}^{\mathrm{SL}}(\tau)$ itself must also be constant along a renormalization group trajectory. This condition dictates that one cannot vary $N_{4}$ while holding fixed the bare couplings $\kappa_{0}$ and $\Delta$ to generate a renormalization group trajectory. Recalling from equation \eqref{renG} that $G_{\mathrm{ren}}\Lambda_{\mathrm{ren}}$ depends on the values of $\bar{s}_{0}$, $\omega_{j}$, and $\langle N_{4}^{(1,4)}\rangle$, all of which depend on the bare couplings, one must make certain that the appropriate combination of $\bar{s}_{0}$, $\omega_{j}$, and $\langle N_{4}^{(1,4)}\rangle$ remains constant as one changes $N_{4}$, $\kappa_{0}$, and $\Delta$. This adjustment potentially leaves some freedom in changes of $N_{4}$, $\kappa_{0}$, and $\Delta$ that preserve the value of $G_{\mathrm{ren}}\Lambda_{\mathrm{ren}}$. One may then bring to bear the general expectations of subsection \ref{guidelines}, possibly using these expectations to reduce or eliminate this residual freedom. The variation of $\tilde{\ell}_{\mathrm{UV}}$, given by the ratio $a/\ell_{\mathrm{P}}$, along a renormalization group trajectory should be in concert with these expectations, with $\tilde{\ell}_{\mathrm{UV}}$ increasing as the renormalization group transformation is iterated. 


\paragraph{Classical scale $\ell_{\mathrm{cl}}$ as the standard unit of length $\ell_{\mathrm{unit}}$}

Consider using the classical scale $\ell_{\mathrm{cl}}$ as the standard unit of length $\ell_{\mathrm{unit}}$. Constancy of the spacetime $4$-volume $V_{4}$, measured in units of $\ell_{\mathrm{cl}}$, now dictates constancy of $\ell_{\mathrm{dS}}/\ell_{\mathrm{cl}}$ upon identification of $V_{4}$ with that of Euclidean de Sitter space. Accordingly, the renormalized cosmological constant $\Lambda_{\mathrm{ren}}$, measured in units of $\ell_{\mathrm{cl}}$, does not flow under renormalization, but the renormalized Newton constant $G_{\mathrm{ren}}$, measured in units of $\ell_{\mathrm{cl}}$ may flow under renormalization. This choice of standard unit of length potentially leaves some freedom in changes of $N_{4}$, $\kappa_{0}$, and $\Delta$ that preserve the value of $\ell_{\mathrm{dS}}/\ell_{\mathrm{cl}}$. One may again bring to bear the general expectations of subsection \ref{guidelines}, possibly using these expectations to reduce or eliminate this residual freedom. The variation of $\tilde{\ell}_{\mathrm{UV}}$, given by the ratio $a/\ell_{\mathrm{cl}}$, along a renormalization group trajectory should be in concert with these expectations, with $\tilde{\ell}_{\mathrm{UV}}$ increasing as the renormalization group transformation is iterated.

This choice of standard unit of length is closest to the original definition of the meter as a standard unit of length---namely as a particular metal bar---in that this bar establishes a scale of classical proportions, neither quantum-mechanical nor cosmological. 

\paragraph{Quantum scale $\ell_{\mathrm{qm}}$ as the standard unit of length $\ell_{\mathrm{unit}}$}

Consider the quantum-mechanical scale $\ell_{\mathrm{qm}}$ as the standard unit of length. Constancy of the spacetime $4$-volume $V_{4}$, measured in units of $\ell_{\mathrm{qm}}$, dictates constancy of the ratio $\ell_{\mathrm{dS}}/\ell_{\mathrm{qm}}$ upon identification of $V_{4}$ with that of Euclidean de Sitter space. Accordingly, the renormalized cosmological constant $\Lambda_{\mathrm{ren}}$, measured in units of $\ell_{\mathrm{qm}}$, does not flow under renormalization, but the renormalized Newton constant $G_{\mathrm{ren}}$, measured in units of $\ell_{\mathrm{qm}}$, may flow under renormalization. This choice of standard unit of length potentially leaves some freedom in changes of $N_{4}$, $\kappa_{0}$, and $\Delta$ that preserve the value of $\ell_{\mathrm{dS}}/\ell_{\mathrm{qm}}$. One may again bring to bear the general expectations of subsection \ref{guidelines}, possibly using these expectations to reduce or eliminate this residual freedom. The variation of $\tilde{\ell}_{\mathrm{UV}}$, given by the ratio $a/\ell_{\mathrm{qm}}$, along a renormalization group trajectory should be in concert with these expectations, with $\tilde{\ell}_{\mathrm{UV}}$ increasing as the renormalization group transformation is iterated.

This choice of standard unit of length is closest to the modern definition of the meter as a standard unit of length---namely in terms of atomic transitions---in that such transitions establish a scale of quantum-mechanical proportions, neither classical nor cosmological. The choice suffers from one obvious drawback: as the scale $\tilde{\ell}_{\mathrm{UV}}$ is incrementally increased, one expects to reach an interval of scales $(\tilde{\ell}_{\mathrm{UV}},\tilde{\ell}_{\mathrm{IR}})$ on which the scale $\ell_{\mathrm{qm}}$ is no longer accessible. 
If one could correlate $\ell_{\mathrm{qm}}$ with $\ell_{\mathrm{cl}}$, as the modern definition of the meter is correlated with the original definition of the meter, then one could use either $\ell_{\mathrm{qm}}$ or $\ell_{\mathrm{cl}}$ as a standard. When $\tilde{\ell}_{\mathrm{UV}}$ has increased sufficiently that $\tilde{\ell}_{\mathrm{qm}}$ is no longer accessible, one could then switch to using $\ell_{\mathrm{cl}}$. The possibility of establishing such a correlation depends on the degree of decoupling between the quantum-mechanical and classical regimes.

\subsubsection{Extraction of the renormalized couplings}

The chosen model has two couplings, the renormalized Newton constant $G_{\mathrm{ren}}$ and the renormalized cosmological constant $\Lambda_{\mathrm{ren}}$. One extracts these couplings from numerical measurements on an ensemble of causal triangulations as explained in subsection \ref{model}. One thus obtains $G_{\mathrm{ren}}$ in units of $a$ as in equation \eqref{renG} and $\Lambda_{\mathrm{ren}}$ in unit of $a$ as in equation \eqref{desitterlengthlatticespacing}. Determining $G_{\mathrm{ren}}/a^{2}$ requires measurements of $N_{4}$, $C_{4}$, $\xi$, $\bar{s}_{0}$, and $\omega_{j}$. Determining $\Lambda_{\mathrm{ren}}a^{2}$ requires measurements of $N_{4}$ and $C_{4}$. The value of $\xi$ stems from measurements of $\langle N_{4}^{(1,4)}\rangle$, $\langle N_{4}^{(2,3)}\rangle$, $\langle N_{4}^{(3,2)}\rangle$, and $\langle N_{4}^{(4,1)}\rangle$, the value of $\bar{s}_{0}$ stems from measurements of $\langle N_{3}^{\mathrm{SL}}(\tau)\rangle$, and the value of $\omega_{j}$ stems from measurements of $\langle n_{3}^{\mathrm{SL}}(\tau)n_{3}^{\mathrm{SL}}(\tau')\rangle$. The first four quantities determine $C_{4}$ and $\xi$. Subsequent fitting of the function \eqref{discreteEdSvolprof} to $\langle N_{3}^{\mathrm{SL}}(\tau)\rangle$ determines $\bar{s}_{0}$. The eigenvalues $\omega_{j}$ follow from diagonalization of $\langle n_{3}^{\mathrm{SL}}(\tau)n_{3}^{\mathrm{SL}}(\tau')\rangle$. 


Alternatively, one can measure $G_{\mathrm{ren}}$ and $\Lambda_{\mathrm{ren}}$ in units $\ell_{\mathrm{dS}}$, $\ell_{\mathrm{P}}$, $\ell_{\mathrm{cl}}$, or $\ell_{\mathrm{qm}}$. 
If one wishes to use either $\ell_{\mathrm{cl}}$ or $\ell_{\mathrm{qm}}$ as the standard unit of length, then one must make numerical measurements to extract these quantities in units of the lattice spacing. Determining $\ell_{\mathrm{cl}}$ and $\ell_{\mathrm{qm}}$ requires measurement of $\langle d_{\mathfrak{s}}(\sigma)\rangle$ followed by the analysis of \cite{JHC2}, which itself calls for the measurement of $\langle N_{4}^{(1,4)}\rangle$, $\langle N_{4}^{(2,3)}\rangle$, $\langle N_{4}^{(3,2)}\rangle$, and $\langle N_{4}^{(4,1)}\rangle$.



\subsubsection{Implementation of a renormalization group transformation}

I choose to achieve indirectly the implementation of a renormalization group transformation. 
To probe different intervals of scales $(\ell_{\mathrm{UV}},\ell_{\mathrm{IR}})$, I propose generating ensembles of causal triangulations characterized by these different intervals of scales. How does one generate an ensemble of causal triangulations characterized by a particular interval of scales? One lets the dynamics do this job. When generating an ensemble, one selects the number $\bar{T}$ of time slices, the number $\bar{N}_{4}$ of $4$-simplices, and the bare couplings $\kappa_{0}$ and $\Delta$ assuming a spacetime manifold of the form $\mathrm{S}^{3}\times\mathrm{S}^{1}$. 
The ensemble average quantum geometry that dynamically results from these chosen values contains information only about a certain interval of scales. How does one then determine this interval of scales? The dimensionless ultraviolet scale $\tilde{\ell}_{\mathrm{UV}}$ is given by the smallest scale present---the lattice spacing $a$---referenced to the chosen standard unit of length $\ell_{\mathrm{unit}}$. 
The dimensionless infrared scale $\tilde{\ell}_{\mathrm{IR}}$ is given by the largest present scale---the de Sitter length $\ell_{\mathrm{dS}}$---referenced to the chosen standard unit of length $\ell_{\mathrm{unit}}$. 

To implement a renormalization group transformation, one thus proceeds as follows. Given an ensemble of causal triangulations characterized by the interval of scales $(\ell_{\mathrm{UV}},\ell_{\mathrm{IR}})$ and certain values for the physical observables $V_{4}$ and $d_{\mathfrak{s}}(u)$, one generates another ensemble of causal triangulations characterized by the interval of scales $(\ell_{\mathrm{UV}}+\delta\ell,\ell_{\mathrm{IR}})$ and the same values for the physical observables $V_{4}$ and $d_{\mathfrak{s}}(u)$. How the interval of scales and the values of the physical observables change with the parameters defining an ensemble is not yet well-established. At least at first, one must work by trial and error.\footnote{One might wonder why I have not developed a coarse graining procedure for the implementation of a renormalization group transformation, especially since this is the standard technique applied to lattice-regularized quantum theories of fields. My motivation is two-fold. First, the coarse graining procedures developed by Johnston, Kownacki, and Krzywicki \cite{DAJ&JPK&AK} and by Catterall, Renken, and Thorleifsson \cite{RLR1,RLR2,GT&SC} for Euclidean dynamical triangulations and even the coarse graining procedure developed by Henson \cite{JH} for causal dynamical triangulations are in fact not well-suited to causal dynamical triangulations: 
the triangulation obtained after just a single iteration of the procedure is no longer necessarily a causal triangulation. (Indeed, the output of Henson's coarse graining procedure might not even be a simplicial manifold.) Since subsequent iterations do not begin with a causal triangulation, the relevance of their outcomes is in doubt. Second, reasoning from the general expectations of subsection \ref{guidelines}, a coarse graining procedure for causal dynamical triangulations should function to reduce simultaneously the number $N_{d+1}$ of $(d+1)$-simplices and the number $T$ of time slices while preserving the causal nature of the triangulation. Undoubtedly, one could design such a coarse graining procedure. Its relevance for ensembles of causal triangulations that one can currently simulate is likely quite limited: owing to the relatively small values of $T_{\mathrm{ph}}$ obtained, very few iterations of the coarse graining procedure could be implemented. 
The method for implementation of renormalization group transformations that I have proposed should allow for much more incremental iterations.}

\subsubsection{Construction of the renormalization group flows}

I construct the renormalization group flows of the gravitational couplings by iterating the renormalization group transformation and extracting the renormalized couplings after each iteration. In practice, one proceeds as follows. One first generates a very large number of ensembles $\mathcal{E}_{\mathrm{S}^{3}}(\bar{T},\bar{N}_{4},\kappa_{0},\Delta)$ of causal triangulations within phase C. One next performs the measurements and analyses described in subsections \ref{observables} and \ref{model} to extract the values of the physical observables selected above. 
One then identifies subsets of these ensembles for which the values of the selected physical observables are all the same. One also identifies the interval of scales $(\ell_{\mathrm{UV}},\ell_{\mathrm{IR}})$ characterizing each ensemble. One finally strings together the ensembles within each subset according to their intervals of scales. This string represents the renormalization group trajectory from which the renormalization group flows of the couplings then follows. One may take $\tilde{\ell}_{\mathrm{UV}}$, given by the ratio $a/\ell_{\mathrm{unit}}$, as the parameter along the renormalization group flow. 


\section{Conclusion}\label{conclusion}

As an approach to the construction of quantum theories of gravity based on a lattice regularization of the path integral, causal dynamical triangulations demands a renormalization group analysis. Based on the general considerations of paper I, 
I here proposed a renormalization group scheme for the causal dynamical triangulations of $(3+1)$-dimensional Einstein gravity. Its aim is the extraction of the renormalization group flows of the couplings of the (hypothetical) continuum limit of the lattice-regularized quantum theory. The nature of these renormalization group flows should allow for the determination of whether or not a continuum limit exists. To understand how such a determination would work, I return to consideration of the phase structure of causal dynamical triangulations. 

Recall from subsection \ref{phases} that the phase diagram of the partition function \eqref{partitionfunctionfixedTN} for the action \eqref{EReggeaction4} 
contains a second order phase transition. The presence of this phase transition raises the prospect of rigorously defining the continuum limit of the quantum theory of gravity so defined. If there exists an ultraviolet fixed point along the phase transition, then one could remove the ultraviolet cutoff at this fixed point. In particular, by tuning the bare couplings to this fixed point, one could keep physical observables finite while letting the number of $(d+1)$-simplices diverge and the lattice spacing vanish. 

Equipped with the renormalization group scheme presented in section \ref{renormalization}, one would attempt to determine if there exists an ultraviolet fixed point as follows. The renormalization group trajectories delineated by the method of subsection \ref{formulation} should exhibit a characteristic behavior in the presence of an ultraviolet fixed point. Moving along a renormalization group trajectory from the interior of phase C to the ultraviolet fixed point on the boundary with phase B, one should find that $N_{4}$ necessarily increases without bound and that $\tilde{\ell}_{\mathrm{UV}}$ necessarily decreases towards zero. That a renormalization group transformation does not effect a change in physics---the condition leading to the delineation of the renormalization group trajectories in the first place---would necessitate these behaviors of $N_{4}$ and $\tilde{\ell}_{\mathrm{UV}}$. On approaching the ultraviolet fixed point, the renormalization group flows of the couplings would dictate the nature of the continuum limit. If all of the couplings parametrizing interactions flow to zero at the ultraviolet fixed point, then the continuum limit is asymptotically free. If a finite number of the couplings parametrizing interactions flow to finite values at the ultraviolet fixed point, and if the rest of the couplings parametrizing interactions flow to zero at the ultraviolet fixed point, then the continuum limit is asymptotically safe. 

This discussion has assumed the existence of an ultraviolet fixed point. If there does not exist an ultraviolet fixed point along the second order phase transition, then renormalization group trajectories from the interior of phase C may not even be connected to the second order phase transition. In this case the lattice-regularized quantum theory only defines an effective description valid on length scales larger than that of the regularization. The renormalization group scheme should nevertheless yield the renormalization group flows of the couplings of this effective description.


Causal dynamical triangulations is not the only approach seeking to define a quantum theory of gravity as a renormalizable quantum theory of fields. The functional renormalization group has emerged as a popular analytic technique complementary to lattice regularization techniques. In this approach one attempts to solve the exact renormalization group equation describing the scale-dependence of the effective (average) gravitational action within some chosen truncation $\mathfrak{t}$ of the space $\mathfrak{T}$ of all theories $\mathscr{T}$. This calculation has been successfully performed for an ever-increasing array of truncations, essentially all of which yield evidence for an asymptotically safe ultraviolet fixed point \cite{MN&MR}. One would eventually want to compare the results of the functional renormalization group approach to those of causal dynamical triangulations. The obvious point of comparison is the nature of the ultraviolet fixed points identified in each approach. Specifically, one would want to compare the critical exponents characterizing each ultraviolet fixed point since the critical exponents determine the universality class of the continuum limit. 

Both the causal dynamical triangulations approach and the functional renormalization group approach aim to achieve nonperturbatively the renormalizability of the quantum theories of gravity that they respectively define. Particularly the advent of Ho\v{r}ava-Lifshitz gravity \cite{PH1} has reignited interest in the prospect of constructing perturbatively renormalizable quantum theories of gravity. There are some indications that the continuum limit of the causal dynamical triangulations of $(3+1)$-dimensional Einstein gravity is some version of quantum Ho\v{r}ava-Lifshitz gravity \cite{CDTandHL}. The recent renormalization group study of Ambj{\o}rn, G\"{o}rlich, Jurkiewicz, Kreienbuehl, and Loll potentially provides further evidence for this possibility \cite{JA&AG&JJ&AK&RL}. If this is indeed the case, then one would expect to find appropriate deviations from the model of subsection \ref{model} as one approaches more closely the ultraviolet fixed point along a given renormalization group trajectory. 
A definitive conclusion awaits a more comprehensive understanding of the physics in the vicinity of the second order phase transition.

\section*{Acknowledgments}

I wish to thank Jan Ambj{\o}rn, Steven Carlip, and Renate Loll for several useful conversations. I also wish to thank Renate Loll for comments on a draft of this paper. I acknowledge support from the Foundation for Fundamental Research on Matter 
itself supported by the Netherlands Organization for Scientific Research. This research was supported in part by the Perimeter Institute for Theoretical Physics. Research at the Perimeter Institute is supported by the Government of Canada through Industry Cananda and by the Province of Ontario through the Ministry of Economic Development and Innovation. This research was also supported in part by the United States Department of Energy under grant DE-FG02-91ER40674 at the University of California, Davis.

\end{document}